\documentclass[11pt]{article}
\pdfoutput=1
\usepackage{jheppub}

\usepackage{amssymb,amsmath}
\usepackage{amsfonts}
\usepackage{mathtools}
\allowdisplaybreaks

\usepackage{physics}
\usepackage{cancel}
\usepackage{enumitem}
\usepackage[usenames,dvipsnames]{xcolor}
\usepackage{bm}
\usepackage{bbm}
\usepackage{slashed}
\usepackage{mathrsfs}

\let\originalleft\left
\let\originalright\right
\renewcommand{\left}{\mathopen{}\mathclose\bgroup\originalleft}
\renewcommand{\right}{\aftergroup\egroup\originalright}

\makeatletter
\newenvironment{equations}[1][]{\subequations\ifx\relax#1\relax\else\label{#1}\fi\align\ignorespaces}{\endalign\ignorespacesafterend\endsubequations}
\def\@spliteq#1{\begin{equation}\begin{split}#1\end{split}\end{equation}}
\def\splitequation{\collect@body\@spliteq}

\makeatother

\newcommand{\diff}{\mathrm{d}}
\newcommand{\de}{\partial}
\newcommand{\vepsilon}{\varepsilon}
\newcommand{\vtheta}{\vartheta}
\newcommand{\vphi}{\varphi}
\newcommand{\vrho}{\varrho}

\newcommand{\0}[1]{\mathring{#1}}

\makeatletter
\@addtoreset{equation}{section}

\makeatother

\newcommand{\nn}{\nonumber}

\makeatletter
\gdef\@fpheader{}
\makeatother

\title{Covariant Fractons and Weitzenb\"ock Torsion}

\author[a,b]{Davide Rovere}

\affiliation[a]{Dipartimento di Fisica e Astronomia ``Galileo Galilei'', Universit\`a di Padova, Via F. Marzolo 8, 35131 Padua, Italy}
\affiliation[b]{INFN, Sezione di Padova, Padua, Italy}

\emailAdd{davide.rovere@studenti.unipd.it}

\abstract{The relation between covariant fracton gauge theory and M{\o}ller-Hayashi-Shirafuji theory of gravity is investigated. The former is the gauge theory of a rank-two symmetric tensor with gauge symmetry given by the double derivative of a scalar parameter; the latter is the most general theory, whose action is quadratic in the Weitzenb\"ock torsion. We show that the solutions of covariant fracton gauge theory describe a subsector of the space of solutions of M{\o}ller-Hayashi-Shirafuji theory, providing a new insight in the relation between covariant fractons and gravity, and elucidating the meaning of covariant fracton theory as a new type of gauge theory.}

\begin{document}
\maketitle

\section{Introduction and outlook}

In this paper we pursue the investigation on covariant fracton gauge theory and its properties, studying the relation between covariant fractons and teleparallel gravity.

Fracton phases, which are excitations characterised by restricted mobility due to the conservation of the dipole moment \cite{Pretko:2017xar, Seiberg:2020bhn, Burnell:2021reh}, were firstly introduced in condensed matter physics \cite{Vijay:2015mka}, but attracted interest in a broader context. In particular, covariant fracton gauge theory, introduced for the first time in \cite{Blasi:2022mbl}, is the gauge theory of a rank-two symmetric tensor $h_{\mu\nu}^{(\text{fr})}$ with gauge symmetry given by the double derivative of a scalar parameter
\begin{equation}\label{GaugeFractonIntro}
\delta\,h^{\text{(fr)}}_{\mu\nu} = \de_\mu\,\de_\nu\,\lambda.
\end{equation}
Covariant fracton gauge theory elucidates the main properties of ordinary fracton excitations, in the context of the higher rank extensions of Maxwell theory \cite{Bertolini:2022ijb}, and it can be thought as an extension of linearised Einstein gravity, invariant with respect to longitudinal diffeomorphisms \cite{Bertolini:2023juh, Afxonidis:2023pdq}. Recently, three-dimensional Chern-Simons-like models with covariant fracton gauge symmetry \cite{Bertolini:2024yur}, and the analogue of the electromagnetic duality in covariant fracton gauge theory \cite{Bertolini:2025jov} have been investigated.

The most general quadratic action in the derivatives of $h_{\mu\nu}^{\text{(fr)}}$, invariant with respect to \eqref{GaugeFractonIntro}, can be written in terms of the two possible scalar contractions of the gauge invariant fracton field strength \cite{Rovere:2024nwc}\footnote{$h_{\mu\nu}^{\text{(fr)}}$ and $f_{\mu\nu\vrho}^{(\text{fr})}$ are respectively denoted with $h_{\mu\nu}$ and $f_{\mu\nu\vrho}$ in \cite{Rovere:2024nwc}. A formally equivalent tensor $f_{\mu\nu\vrho}$ appeared previously in \cite{Deser:2006zx}, where the theory of a partially massless spin-two field in de Sitter space is investigated. It appears also in \cite{Bertolini:2024apg} at the linear level.}
\begin{equation}
f^{\text{(fr)}}_{\mu\nu\vrho} = \de_\mu\,h_{\nu\vrho}^{\text{(fr)}}-\de_\nu\,h_{\mu\vrho}^{\text{(fr)}},
\end{equation}
which is analogous to the field strength in electromagnetism. The action reads
\begin{equation}\label{FractonAction}
S_{\text{fr}} =\int\diff^d x\,(\alpha\,{f^{\text{(fr)}}}^{\mu\nu\vrho}\,f^{\text{(fr)}}_{\mu\nu\vrho} +\beta\,{f^{\text{(fr)}}}^{\mu\nu}{}_\nu\,{f^{\text{(fr)}}}_{\mu\vrho}{}^\vrho),
\end{equation}
where $\alpha$ and $\beta$ are arbitrary constants. Taking $\alpha$ and $\beta$ such that $\frac{\beta}{\alpha} = -2$, and replacing the fracton gauge field with the perturbation of the metric in linearised gravity, one gets the Fierz-Pauli action (see equation \eqref{FP}) \cite{Fierz:1939ix}, which is invariant under the full group of linearised diffeomorphisms.

Since the whole dynamics of covariant fracton gauge theory is encoded in the fracton field strength, one might ask which its  gravitational analogue is. We observe in this paper that the covariant fracton field strength is the analogue of the linearised \emph{Weitzenb\"ock torsion} in gravity, when the linearised vielbein has vanishing totally antisymmetric part. The first attempts of a torsion formulation of General Relativity, usually known as \emph{teleparallel equivalent of General Relativity}, date back to Einstein \cite{Einstein1928}, who unsuccessfully tried to unificate gravitation and electromagnetism in four dimensions, using the sixteen degrees of freedom of the vierbein, according to Cartan's \cite{Cartan1922, Cartan1923, Cartan1924, Cartan1925} and Weitzenb\"ock's \cite{Weitzenbock1928} works. For the history and prehistory of the teleparallel gravity see for example \cite{Shirafuji:1995xc, Maluf:2013gaa}. In teleparallel gravity the dynamical variables are encoded in the torsion of a flat connection, known as \emph{Weitzenb\"ock connection}:
\begin{equation}
W_\mu{}^\vrho{}_\nu = e^\vrho{}_a\,\de_\mu\,e_\nu{}^a.
\end{equation}
where $e_\mu{}^a$ is the vielbein, and $e^\mu{}_a$ its inverse. The Weitzenb\"ock torsion, which is the antisymmetric part of the previous connection
\begin{equation}
T_\mu{}^\vrho{}_\nu = W_{[\mu}{}^\vrho{}_{\nu]} = e^\vrho{}_a\,\de_{[\mu}\,e_{\nu]}{}^a,
\end{equation} 
is a well-defined covariant tensor, which is very useful for its geometrical implications: namely, it was employed in the geometrisation of gauge symmetries in the context of Scherk-Schwarz compactifications \cite{Scherk:1979zr}, and in generalised geometry \cite{Coimbra:2011ky, Aldazabal:2013mya, Lee:2014mla}. M{\o}ller studied extensions of General Relativity in teleparallel formulation, by considering the wider class of theories defined by the most general action quadratic in the Weitzenb\"ock torsion \cite{Moller1961a, Moller1961b, Moller1978}. At the same time, independently, the possible phenomenological implications of the model were first studied by Hayashi and Shirafuji \cite{Hayashi:1979qx}. The model can be also viewed as the gauge theory of translations \cite{Hayashi:1967se}. The action for the theory considered by M{\o}ller and by Hayashi and Shirafuji, which we will call \emph{M{\o}ller-Hayashi-Shirafuji theory} (\textsc{mhs}), reads
\begin{equation}\label{ActionMHSIntro}
S_{\textsc{mhs}} = \frac{1}{2}\int \diff^d x\,e\,(\alpha_1\,T^{\mu\nu\vrho}\,T_{\mu\nu\vrho} + \alpha_2\,T^{\mu\nu\vrho}\,T_{\mu\vrho\nu} + \alpha_3\,T^{\mu\nu}{}_\nu\,T_{\mu\vrho}{}^\vrho),
\end{equation}
where $\alpha_1, \alpha_2, \alpha_3$ are arbitrary constants. 

We will show that, when $\frac{\beta}{\alpha} \neq 2$, the solutions of the equations of motion of covariant fracton gauge theory can be viewed as the subsector in the space of solutions of the equations of motion of linearised \textsc{mhs} theory, identified by the vanishing antisymmetric part of the linearised vielbein, denoted with $b_{\mu\nu}$,
\begin{equation}\label{bvanishing}
b_{\mu\nu} = 0.
\end{equation}
The embedding of covariant fracton solutions in \textsc{mhs} theory is realised when the fracton gauge field $h_{\mu\nu}^{(\text{fr})}$ is identified with the symmetric part of the linearised vielbein, denoted with $h_{\mu\nu}$,
\begin{equation}
h_{\mu\nu}^{\text{(fr)}} = h_{\mu\nu},
\end{equation} 
and the constants $\alpha$ and $\beta$ in the covariant fracton action \eqref{FractonAction} are identified with combinations of the constant $\alpha_1, \alpha_2$, and $\alpha_3$ in the \textsc{mhs} action \eqref{ActionMHSIntro}
\begin{equation}
\alpha = \tfrac{1}{4}\,(2\,\alpha_1 + \alpha_2), \quad
\beta = \tfrac{1}{2}\,\alpha_3.
\end{equation} 
To show this, focusing on four dimensions for definiteness, we will compare the particle spectrum of both theories. Except for some particular cases, tuned by special values of the constants $\alpha_1, \alpha_2$, and $\alpha_3$, such that the gauge invariance is enlarged, the spectrum of \textsc{mhs} theory in four dimensions consists in six degrees of freedom, given by a particle with helicity $\pm 2$, one with helicity $\pm 1$, and two scalar particles. The antisymmetric field $b_{\mu\nu}$  describes only a scalar particle (as expected, since a two-form gauge field in four dimensions is dual to a scalar field). The sector with vanishing $b_{\mu\nu}$ coincides with the particle spectrum of covariant fracton gauge theory, which indeed describes the remaining five propagating degrees of freedom \cite{Afxonidis:2023pdq}, except for the particular cases in which the gauge symmetry is enlarged. 

Linearised diffeomorphisms act on $b_{\mu\nu}$ as a one-form gauge transformation for a two-form field (usually known as Ramond-Kalb field \cite{Kalb:1974yc}). The condition \eqref{bvanishing} is not diffeomorphism invariant, but it leaves a residual gauge symmetry  when the gauge parameter is exact, that is, when it is a derivative. This is the case of longitudinal diffeomorphisms, so that the residual gauge symmetry is the fracton gauge symmetry. In a sense, covariant fracton solutions spontaneously break the diffeomorphism symmetry of \textsc{mhs} theory, preserving only the fracton gauge symmetry. 

To elucidate this setting, we will take into account the \textsc{brst} formulation \cite{Becchi:1974md, Becchi:1974xu, Becchi:1975nq, Tyutin:1975qk} of covariant fracton gauge theory. In the \textsc{brst} setup, the gauge parameter is replaced by an anticommuting scalar field (fracton ghost) $\lambda$, and the gauge transformation is realised by a nilpotent differential operator $s$ on the space of the fracton gauge field and of the ghost and their derivatives, evaluated in the same spacetime point. The local $s$-cohomology, which classifies the consistent anomalies, is generated by the undifferentiated ghost and by the fracton field strength and its derivatives, as in the analogous case of electromagnetism \cite{Rovere:2024nwc}. 

We will show that the \textsc{brst} transformations of covariant fracton gauge theory can be conveniently formulated in an equivalent way, by including two \textsc{brst} trivial doublets. In this formulation a symmetric rank-two tensor $h_{\mu\nu}$ and an antisymmetric one $b_{\mu\nu}$ are included. They can be respectively viewed as the symmetric and the antisymmetric part of the linearised vielbein, but the \textsc{brst} transformations do not involve any local Lorentz transformation: this is the case of Weitzenb\"ock geometry, which is flat and it is invariant only under global Lorentz transformations \cite{Hayashi:1977jd}. Nevertheless, on the one side, the \textsc{brst} transformations require the restriction of linearised diffeomorphisms to the ``longitudinal" ones, bringing us to covariant fracton gauge transformations as they were first introduced \cite{Blasi:2022mbl}; on the other side, $b_{\mu\nu}$ is equivalent to zero.  
This individuates the subsector of \textsc{mhs} theory in which covariant fracton gauge theory lives. 
 
Covariant fractons were introduced in \cite{Blasi:2022mbl} only at linear level. In \cite{Afxonidis:2024tph} it is shown that the theory is afflicted by instabilities, which in principle can be cured by considering suitable interactions. Going towards this direction, the result in this note can be viewed as a possible way to embed covariant fracton theory in a larger theory of gravity, whose non-linear extension is automatically given by the extension of General Relativity in teleparallel formulation, studied by M{\o}ller, Hayashi, and Shirafuji. 

A possible non-linear extension of covariant fracton gauge theory is also discussed in \cite{Bertolini:2024apg}, where the Fr\"olicher-Nijenhuis bracket is employed, acting on vector-valued differential forms and coinciding with the Lie derivative between vector fields when it acts on vectors \cite{FroNij1, FroNij2, FroNij3}. The non-linear extension of the fracton gauge transformation is realised by using a vector parameter, which is restricted to be the derivative of a scalar parameter only at the linear level; the gauge field is replaced by a vector-valued one-form, and a constraint is added in order to make it symmetric; a possible non-linear equation of motion is postulated, without an action from which to deduce it; the linearisation of such an equation is the equation of motion of covariant fracton gauge theory (with $\beta = 0$ in \eqref{FractonAction}).

It would be interesting to exploit the relation between covariant fractons and Weitzenb\"ock torsion in understanding deeper this non-linear extension. In particular, one could try to see if it is possible to relate the postulated non-linear equation of motion with the equations of \textsc{mhs} theory, since it seems natural to identity the vector-valued one form with the linearised vielbein.

The rest of the paper is organised as follows: in Section \ref{FractonTorsion} covariant fracton theory is formulated in a way which makes manifest the relation with torsion formulation of gravity; in Section \ref{GenRelTorsion}, \ref{MHSTorsion}, and \ref{LinTorsion} torsion formulation of General Relativity, the \textsc{mhs} theory, and its linearisation are reviewed; Section \ref{ParticleFracton} and \ref{ParticleMHS} the particle spectrum of covariant fracton and \textsc{mhs} theory are computed and compared, providing the embedding of covariant fractons in \textsc{mhs} theory; finally, in Section \ref{BRSTorsion} the equivalent \textsc{brst} formulation is investigated. In Appendix \ref{SpinConnection} there is a summary of Weitzenb\"ock geometry; in Appendix \ref{BRSCohomology} a computation for \textsc{brst} cohomology is shown; Appendix \ref{Conventions} is devoted to a comparison between different conventions; and in Appendix \ref{polarisation} there is a summary of polarisation matrices.

\section{Covariant fracton gauge theory in ``torsion-like formulation"}\label{FractonTorsion}

In this section covariant fracton gauge theory \cite{Blasi:2022mbl} is formulated in a way in which its embedding in gravitational theories in torsion formulation will be natural. 

Covariant fracton gauge theory is defined by a symmetric rank-two tensor $h^{(\text{fr})}_{\mu\nu}$ and by a scalar gauge parameter, promoted to an anticommuting ghost $\lambda$ in \textsc{brst} formalism \cite{Becchi:1974md, Becchi:1974xu, Becchi:1975nq, Tyutin:1975qk}. The nilpotent \textsc{brst} transformations are  
\begin{equation}\label{BRSFracton}
s\,h^{(\text{fr})}_{\mu\nu} = \de_\mu\,\de_\nu\,\lambda,  \quad
s\,\lambda = 0.
\end{equation}
One defines the \emph{fracton field strength} 
\begin{equation}
f^{(\text{fr})}_{\mu\nu\vrho} := \de_\mu\,h^{(\text{fr})}_{\nu\vrho} - \de_\nu\,h^{(\text{fr})}_{\mu\vrho},
\end{equation}
which is antisymmetric in the first two indices, and whose totally antisymmetric part vanishes (algebraic Bianchi identity). Moreover, the fracton field strength is \textsc{brst} invariant and it satisfies a differential Bianchi identity:
\begin{equations}
& s\,f^{\text{(fr)}}_{\mu\nu\vrho} = 0,\\
& f^{\text{(fr)}}_{\mu\nu\vrho} + 
f^{\text{(fr)}}_{\nu\vrho\mu} +
f^{\text{(fr)}}_{\vrho\mu\nu} = 0,\label{BianchiIdAlg}\\
& \de_\sigma\,f^{\text{(fr)}}_{\mu\nu\vrho} + \de_\mu\,f^{\text{(fr)}}_{\nu\sigma\vrho} + \de_\nu\,f^{\text{(fr)}}_{\sigma\mu\vrho} = 0.\label{BianchiIdDiff}
\end{equations}
The most general invariant quadratic action in the derivative of $h^{(\text{fr})}_{\mu\nu}$ can be written in terms of the fracton field strength. There are only two independent quadratic scalar contractions of the fracton field strength, since it transforms in the mixed Young tableau $(2,1)$.\footnote{\label{Young} We denote with $(\lambda_1,\dots,\lambda_n)$ the Young tableau with $n$ rows and $\lambda_i$ boxes in the $i$th row, $i=1,\dots,n$. $(\lambda_1,\dots,\lambda_n)_t$ is the corresponding traceless part.} So, the most general quadratic invariant action can be written as
\begin{align}\label{ActionFr}
S_{\text{fr}} &=\int\diff^d x\,(\alpha\,{f^{\text{(fr)}}}^{\mu\nu\vrho}\,f^{\text{(fr)}}_{\mu\nu\vrho} +\beta\,{f^{\text{(fr)}}}^{\mu\nu}{}_\nu\,{f^{\text{(fr)}}}_{\mu\vrho}{}^\vrho) = \nn\\
&= \int\diff^d x\,(2\,\alpha\,\de^\vrho\,h^{\text{(fr)}\,\mu\nu}\,\de_\vrho\,h^{\text{(fr)}}_{\mu\nu} - 2\,(\alpha-\tfrac{\beta}{2})\,\de^\nu\,h^{\text{(fr)}}_{\mu\nu}\,\de_\vrho\,h^{\text{(fr)}\,\mu\vrho} \,+\nn\\
& \qquad + \beta\,\de_\mu\,h^{\text{(fr)}\,\nu}{}_\nu\,\de^\mu\,h^{\text{(fr)}\,\vrho}{}_\vrho - 2\,\beta\,\de_\mu\,h^{\text{(fr)}\,\nu}{}_\nu\,\de_\vrho\,h^{\text{(fr)}\,\mu\vrho}),
\end{align}
for arbitrary constants $\alpha$ and $\beta$. The corresponding equations of motion require the following symmetric tensor to vanish
\begin{align}
E^{\text{(fr)}}_{\mu\nu} := -\frac{1}{2}\,\frac{\delta\,S_{\text{fr}}}{\delta\,{h^{(\text{fr})}}^{\mu\nu}} &= 
\alpha\,\de_\vrho\,{f^{\text{(fr)}}}^\vrho{}_{(\mu\nu)} 
-\tfrac{\beta}{2}\,\de_{(\mu}\,{f^{\text{(fr)}}}_{\nu)\vrho}{}^\vrho 
+\beta\,\eta_{\mu\nu}\,\de_\vrho\,{f^{\text{(fr)}}}^{\vrho\sigma}{}_\sigma = \label{EqFractonf}\\
&= 2\,\alpha\,\de^2\,h_{\mu\nu}^{\text{(fr)}} - \tfrac{2\,\alpha-\beta}{2}\,\de_{(\mu}\,\de^\vrho\,h^{\text{(fr)}}_{\nu)\vrho} -\beta\,\de_\mu\,\de_\nu\,h^{\text{(fr)}\,\vrho}{}_\vrho \,+\nn\\
&\quad + \beta\,\eta_{\mu\nu}\,(\de^2\,h^{\text{(fr)}\,\vrho}{}_\vrho-\de_\vrho\,\de_\sigma\,h^{\text{(fr)}\,\vrho\sigma})
\end{align}
where the round brackets among indices denote symmetrisation \emph{without} any numerical coefficient.\footnote{For example, $A_{(\mu\nu)}=A_{\mu\nu}+A_{\nu\mu}$. Similarly, $A_{[\mu\nu]}=A_{\mu\nu}-A_{\nu\mu}$.} It is useful to define the following tensor
\begin{align}
R^{\text{(fr)}}_{\mu\nu\vrho\sigma} &= 
\tfrac{1}{2}\,\de_\sigma\,\de_\mu\,h^{(\text{fr})}_{\nu\vrho}
-\tfrac{1}{2}\,\de_\sigma\,\de_\nu\,h^{(\text{fr})}_{\mu\vrho}
-\tfrac{1}{2}\,\de_\vrho\,\de_\mu\,h^{(\text{fr})}_{\nu\sigma}
+\tfrac{1}{2}\,\de_\vrho\,\de_\nu\,h^{(\text{fr})}_{\mu\sigma} =\nn\\
&= \tfrac{1}{2}\,\de_\sigma\,f^{\text{(fr)}}_{\mu\nu\vrho} -\tfrac{1}{2}\,\de_\vrho\,f^{\text{(fr)}}_{\mu\nu\sigma}, 
\end{align}
which enjoys the same symmetries as the Riemann tensor, being antisymmetric in $\mu\nu$ and in $\vrho\sigma$ and symmetric in $\mu\nu\leftrightarrow\vrho\sigma$. The corresponding ``Ricci tensor" and ``Ricci scalar" respectively read\footnote{Consider that $ \de_{[\mu}\,{f^{\text{(fr)}}}_{\nu]\vrho}{}^\vrho - \de_\vrho\,{f^{\text{(fr)}}}^\vrho{}_{[\mu\nu]} = 0$ as a consequence of \eqref{BianchiIdAlg}--\eqref{BianchiIdDiff}.}
\begin{align}
R^{(\text{fr})}_{\mu\nu} &= {R^{\text{(fr)}}}^\vrho{}_{\mu\vrho\nu} = -\tfrac{1}{4}\,\de_\vrho\,{f^{\text{(fr)}}}^\vrho{}_{(\mu\nu)} - \tfrac{1}{4}\,\de_{(\mu}\,{f^{\text{(fr)}}}_{\nu)\vrho}{}^\vrho,\label{Ricci}\\
R^{(\text{fr})} &= {R^{(\text{fr})}}_\mu{}^\mu = -\de_\mu\,{f^{\text{(fr)}}}^{\mu\nu}{}_\nu. \label{RicciScalar}
\end{align}
Writing $\de_{(\mu}\,{f^{\text{(fr)}}}_{\nu)\vrho}{}^\vrho$  and $\de_\mu\,{f^{\text{(fr)}}}^{\mu\nu}{}_\nu$ in terms of the ``Ricci tensor" and of the ``Ricci scalar" in the equations of motion \eqref{EqFractonf}, one recognises the analogue of the linearised Einstein tensor: 
\begin{equation}\label{EqFractonf2}
E^{\text{(fr)}}_{\mu\nu} = 
2\,\beta\,(R^{\text{(fr)}}_{\mu\nu} - \tfrac{1}{2}\,R^{\text{(fr)}}\,\eta_{\mu\nu})
+(\alpha + \tfrac{\beta}{2})\,\de_\vrho\,{f^{\text{(fr)}}}^\vrho{}_{(\mu\nu)}.
\end{equation}
Indeed, if one replace $h^{\text{(fr)}}_{\mu\nu}$ in the action \eqref{ActionFr} with a rank-two symmetric tensor $h_{\mu\nu}$, thought as the perturbation of a metric $g_{\mu\nu} = \eta_{\mu\nu} + h_{\mu\nu}$, and $\alpha$ and $\beta$ are such that
\begin{equation}\label{GenRelCond1}
\tfrac{\beta}{\alpha} = -2,
\end{equation}
choosing the normalisation $\alpha = \frac{1}{2}$, one recovers the massless Fierz-Pauli action \cite{Fierz:1939ix}
\begin{equation}\label{FP}
S_{\textsc{fp}} = \tfrac{1}{2}\int\diff^d x\,(\de_\vrho\,h_{\mu\nu}\,\de^\vrho\,h^{\mu\nu} - 2\,\de^\nu\,h_{\mu\nu}\,\de_\vrho\,h^{\mu\vrho} + 2\,\de_\mu\,h_\nu{}^\nu\,\de_\vrho\,h^{\mu\vrho} - \de_\mu\,h_\nu{}^\nu\,\de^\mu\,h_\vrho{}^\vrho).
\end{equation}
Consistently, imposing the condition \eqref{GenRelCond1} in the equations of motion \eqref{EqFractonf2}, the last term vanishes and one gets the linearised Einstein equations.

Using the equations of motion in the form \eqref{EqFractonf2}, it is manifest that the first divergence is not identically zero, unless in the Fierz-Pauli case, since only the analogue of the Einstein tensor is transverse:
\begin{equation}\label{div}
\de^\mu\,E^{\text{(fr)}}_{\mu\nu} = (\alpha + \tfrac{\beta}{2})\,\de^\mu\,\de_\vrho\,{f^{\text{(fr)}}}^\vrho{}_{(\mu\nu)}.
\end{equation}
Instead, one gets a transversality condition for all values of $\alpha$ and $\beta$ by taking the second divergence, as a consequence of the invariance under the fracton gauge transformations \eqref{BRSFracton},
\begin{equation}
\de^\mu\,\de^\nu\,E^{\text{(fr)}}_{\mu\nu} = 
(\alpha + \tfrac{\beta}{2})\,\de^\mu\,\de^\nu\,\de^\vrho\,f^{\text{(fr)}}_{\mu\nu\vrho} = 0,
\end{equation}
which vanishes since the fracton field strength is antisymmetric in its first two indices.

\section{Review of teleparallel gravity}\label{GenRelTorsion}

In the usual formulation of General Relativity, one chooses a connection $\Gamma_\mu{}^\vrho{}_\nu$ compatible with the metric $g_{\mu\nu}$, such that the torsion vanishes and the curvature encodes the dynamical variables:
\begin{equations}
\nabla_\vrho(\Gamma)\,g_{\mu\nu} &= \de_\vrho\,g_{\mu\nu} - \Gamma_\vrho{}^\sigma{}_\mu\,g_{\sigma\nu} - \Gamma_\vrho{}^\sigma{}_\nu\,g_{\mu\sigma} = 0, \\
T_\mu{}^\vrho{}_\nu(\Gamma) &= \Gamma_\mu{}^\vrho{}_\nu - \Gamma_\nu{}^\vrho{}_\mu = 0,\\
R_{\mu\nu}{}^\vrho{}_\sigma(\Gamma) &= \de_\mu\,\Gamma_\nu{}^\vrho{}_\sigma - \de_\nu\,\Gamma_\mu{}^\vrho{}_\sigma + \Gamma_\mu{}^\vrho{}_\tau\,\Gamma_\nu{}^\tau{}_\sigma - \Gamma_\nu{}^\vrho{}_\tau\,\Gamma_\mu{}^\tau{}_\sigma.
\end{equations}
As known, there is a unique choice for such a connection, the Levi-Civita connection: 
\begin{equation}
\Gamma_\mu{}^\vrho{}_\nu = \tfrac{1}{2}\,g^{\vrho\sigma}\,(\de_\mu\,g_{\nu\sigma} + \de_\nu\,g_{\mu\sigma} - \de_\sigma\,g_{\mu\nu}).
\end{equation}
In the following we will denote with $\nabla_\mu$ the covariant derivative with respect to the Levi-Civita connection $\Gamma_\mu{}^\vrho{}_\nu$, and with $R^\vrho{}_{\sigma\mu\nu}$ the Riemann tensor.

There is an alternative but equivalent formulation of General Relativity (at least classically and without boundary terms), known as \emph{teleparallel equivalent of General Relativity}, in which one chooses a connection $W_\mu{}^\vrho{}_\nu$, compatible with the metric, such that the dynamical variables are encoded in the torsion and the curvature vanishes \cite{Einstein1928, Weitzenbock1928}:
\begin{equations}
\nabla_\vrho(W)\,g_{\mu\nu} &= \de_\vrho\,g_{\mu\nu} - W_\vrho{}^\sigma{}_\mu\,g_{\sigma\nu} - W_\vrho{}^\sigma{}_\nu\,g_{\mu\sigma} = 0, \label{WMetricCompatible}\\
T_\mu{}^\vrho{}_\nu(W) &= W_\mu{}^\vrho{}_\nu - W_\nu{}^\vrho{}_\mu,\\
R_{\mu\nu}{}^\vrho{}_\sigma(W) &= \de_\mu\,W_\nu{}^\vrho{}_\sigma - \de_\nu\,W_\mu{}^\vrho{}_\sigma + W_\mu{}^\vrho{}_\tau\,W_\nu{}^\tau{}_\sigma - W_\nu{}^\vrho{}_\tau\,W_\mu{}^\tau{}_\sigma = 0.
\end{equations}
While in the usual formulation of General Relativity the gravitational interaction is due to the curvature of spacetime, in teleparallel formulation the gravitational interaction is mediated by a Lorentz-like force in a flat spacetime, as in Yang-Mills theories \cite{deAndrade:1997gka}. Indeed, while in the former case the action is linear in the curvature, in the latter one the action is quadratic in the torsion. 

Denoting with $e_\mu{}_a$ the components of the vielbein and with $e^\mu{}_a$ the components of the frame or inverse vielbein, it is possible to choose a connection $W_\mu{}^\vrho{}_\nu$ such that the vielbein is parallel or covariantly constant, that is the following equation holds:
\begin{equation}\label{parallel}
\de_\mu\,e_\nu{}^a - W_\mu{}^\vrho{}_\nu\,e_\vrho{}^a = 0.
\end{equation}
The unique connection solving the previous condition is the \emph{Weitzenb\"ock connection} \cite{Weitzenbock1928}
\begin{equation}
W_\mu{}^\vrho{}_\nu = e^\vrho{}_a\,\de_\mu\,e_\nu{}^a,
\end{equation}
which consistently has vanishing curvature, since it is the 
Cartan-Maurer form out of the vielbein.\footnote{The spin connection corresponding to the Weitzenb\"ock connection is zero (\emph{Weitzenb\"ock condition}). See Appendix \ref{SpinConnection} for more details.} Moreover, the condition \eqref{parallel} implies the compatibility condition \eqref{WMetricCompatible}, since
\begin{equation}
g_{\mu\nu} = e_\mu{}^a\,e_\nu{}^b\,\eta_{ab}.
\end{equation}
In the following we will denote with $D_\mu$ the covariant derivative with respect to the Weitzenb\"ock connection $W_\mu{}^\vrho{}_\nu$, and with $T_\mu{}^\vrho{}_\nu$ the Weitzenb\"ock torsion, 
\begin{equation}
T_\mu{}^\vrho{}_\nu = e^\vrho{}_a\,\de_{[\mu}\,e_{\nu]}{}^a,
\end{equation}
antisymmetric in the first and in the last index. 

The Weitzenb\"ock torsion satisfies the following Bianchi identity
\begin{equation}\label{BianchiWeitzenbock}
D_{[\mu}\,T_{\nu]}{}^\sigma{}_\vrho + D_\vrho\,T_\mu{}^\sigma{}_\nu -T_{[\mu}{}^{\sigma\tau}\,T_{\nu]\tau\vrho} -T_{\mu\tau\nu}\,T_\vrho{}^{\sigma\tau} = 0,
\end{equation}
whose trace, which will be useful in the following, reads
\begin{equation}\label{TraceBianchiWeitzenbock}
D_{[\mu}\,T_{\nu]\vrho}{}^\vrho + D_\vrho\,T_\mu{}^\vrho{}_\nu 
+ T_{\mu\vrho\nu}\,T^{\vrho\sigma}{}_\sigma = 0.
\end{equation}
Using the derivative of the vielbein determinant $\de_\mu\,e = e\,W_\mu{}^\vrho{}_\vrho$, one can show that the following identity holds for the Weitzenb\"ock covariant divergence, for any vector $V^\mu$,
\begin{equation}\label{CovariantWeitzenbockDivergence}
e\,D_\mu\,V^\mu = \de_\mu\,(e\,V^\mu) + e\,T_\mu{}^\mu{}_\nu\,V^\nu.
\end{equation}

The relation between the Levi-Civita connection and the Weitzenb\"ock is given by:
\begin{equation}\label{Levi-Civita-Weitzenbock}
\Gamma_\mu{}^\vrho{}_\nu = W_\mu{}^\vrho{}_\nu + K_\mu{}^\vrho{}_\nu = W_\mu{}^\vrho{}_\nu + \tfrac{1}{2}\,(T_{\mu\nu}{}^\vrho + T_{\nu\mu}{}^\vrho - T_\mu{}^\vrho{}_\nu),
\end{equation}
where the tensor $-K_\mu{}^\vrho{}_\nu$ is the \emph{contorsion}.
Using the vanishing of the curvature of the Weitzenb\"ock connection and the relation \eqref{Levi-Civita-Weitzenbock}, one can write the Riemann tensor in terms of the contorsion in the following way:
\begin{equation}
R^\mu{}_{\nu\vrho\sigma} = D_\vrho\,K_\sigma{}^\mu{}_\nu - D_\sigma\,K_\rho{}^\mu{}_\nu + T_\vrho{}^\tau{}_\sigma\,K_\tau{}^\mu{}_\nu + K_\vrho{}^\mu{}_\tau\,K_\sigma{}^\tau{}_\nu - K_\sigma{}^\mu{}_\tau\,K_\vrho{}^\tau{}_\nu.
\end{equation}
Therefore, using the trace of the Bianchi identity \eqref{TraceBianchiWeitzenbock}, the Ricci tensor is
\begin{align}
R_{\mu\nu} = R^\vrho{}_{\mu\vrho\nu} &= \tfrac{1}{2}\,D_\vrho\,T_{(\mu\nu)}{}^\vrho - \tfrac{1}{2}\,D_{(\mu}\,T_{\nu)\vrho}{}^\vrho + \tfrac{1}{2}\,T_{(\mu\nu)\vrho}\,T^{\vrho\sigma}{}_\sigma \,+\nn\\
& + \tfrac{1}{4}\,T_{\vrho\mu\sigma}\,T^\vrho{}_\nu{}^\sigma - \tfrac{1}{2}\,T_{\mu\vrho\sigma}\,T_{\nu}{}^{\vrho\sigma} - \tfrac{1}{2}\,T_{\mu\vrho\sigma}\,T_{\nu}{}^{\sigma\vrho},
\label{RicciWeitzenbock}
\end{align}
and the Ricci scalar is 
\begin{equation}\label{RicciScalarWeitzenbock}
R = R^\mu{}_\mu = -2\,D_\mu\,T^{\mu\nu}{}_\nu  - \tfrac{1}{4}\,T^{\mu\nu\vrho}\,T_{\mu\nu\vrho} - \tfrac{1}{2}\,T^{\mu\nu\vrho}\,T_{\mu\vrho\nu} - T_{\mu\nu}{}^\nu\,T^{\mu\vrho}{}_\vrho.
\end{equation}
Finally, using the Weitzenb\"ock covariant divergence \eqref{CovariantWeitzenbockDivergence}, one gets  the following identity
\begin{equation}
e\,R = -\de_\mu\,(2\,e\,T^{\mu\nu}{}_\nu) - e\,(\tfrac{1}{4}\,T^{\mu\nu\vrho}\,T_{\mu\nu\vrho} + \tfrac{1}{2}\,T^{\mu\nu\vrho}\,T_{\mu\vrho\nu} - T_{\mu\nu}{}^\nu\,T^{\mu\vrho}{}_\vrho),
\end{equation}
which shows that, up to a boundary term, the Einstein-Hilbert action admits an equivalent formulation in terms of a linear combination of the quadratic scalar contractions of the Weitzenb\"ock torsion \cite{Moller1961a, Moller1961b, Moller1978, Hayashi:1967se, Hayashi:1979qx, deAndrade:1997gka}:
\begin{equation}\label{HE}
S_{\textsc{eh}} = -\frac{1}{2}\int\diff^d x\,e\,R = \frac{1}{2}\int\diff^d x\,e\,\left(\frac{1}{4}\,T^{\mu\nu\vrho}\,T_{\mu\nu\vrho} + \frac{1}{2}\,T^{\mu\nu\vrho}\,T_{\mu\vrho\nu} - T_{\mu\nu}{}^\nu\,T^{\mu\vrho}{}_\vrho\right),
\end{equation}
where we set $\kappa = 8\,\pi\,G = 1$.\footnote{This proves the dynamical equivalence of the usual formulation of General Relativity and of the teleparallel equivalent, since it is  not necessary to assume the existence of a globally defined frame \cite{Cederwall:2021xqi}. Although not manifest in teleparallel formulation, the precise combination in \eqref{HE} is locally Lorentz invariant (see next Sections and \cite{Cederwall:2021xqi}), and the inverse frame serves as a gauge field. Thus, the global extension of the frame is ensured by suitably choosing the gauge redundancy in the overlaps of the coordinates patches.}

\section{Review of M{\o}ller-Hayashi-Shirafuji theory}\label{MHSTorsion}

The relation \eqref{HE} shows that the Einstein-Hilbert action consists in a linear combination of three quadratic scalar contractions of the Weitzenb\"ock torsion with precise relative coefficients. In \cite{Moller1961a, Moller1961b, Moller1978, Hayashi:1967se, Hayashi:1979qx} the theory defined by the same action with arbitrary coefficients is considered,
\begin{equation}\label{MollerAction}
S_{\textsc{mhs}} = \frac{1}{2}\int \diff^d x\,e\,(\alpha_1\,T^{\mu\nu\vrho}\,T_{\mu\nu\vrho} + \alpha_2\,T^{\mu\nu\vrho}\,T_{\mu\vrho\nu} + \alpha_3\,T^{\mu\nu}{}_\nu\,T_{\mu\vrho}{}^\vrho).
\end{equation}
We will call the theory described by this action \emph{M{\o}ller-Hayashi-Shirafuji theory} (\textsc{mhs}).\footnote{Hayashi and Shirafuji called it \emph{New General Relativity} \cite{Hayashi:1979qx}. See Appendix \ref{Conventions} for a comparison with definitions and conventions in this paper and in \cite{Moller1978} and in \cite{Hayashi:1979qx}.} 

The peculiar combination picking the case of General Relativity, which is 
\begin{equation}\label{GenRelCond2}
\alpha_1 = \tfrac{1}{4}, \quad 
\alpha_2 = \tfrac{1}{2}, \quad
\alpha_3 = -1,
\end{equation}
is captured by formally requiring the local Lorentz invariance, which indeed is the necessary requirement in the vielbein formulation of General Relativity, beside diffeomorphism invariance, in order to recover the right number of degrees of freedom \cite{Cho:1975dh}.\footnote{Since in \textsc{mhs} theory local Lorentz-invariance is spoiled, unless \eqref{GenRelCond2} are met, one has to assume the parallelisability of spacetime a priori, that is the existence of a frame which serves as basis for the tangent bundle, which trivialises \cite{Hayashi:1979qx}.}

The \textsc{mhs} action is the most general quadratic action in the Weitzenb\"ock torsion, since there are only three possible independent quadratic scalar contractions of the torsion. Indeed, the torsion has three irreducible components:\footnote{See note \ref{Young} for the notation of Young tableaux.}
\begin{equation}
T_{\mu\vrho\nu} \rightarrow (1,1) \otimes (1) = (1,1,1) \otimes (2,1)_t \otimes (1).
\end{equation}
$(1,1,1)$ is the totally antisymmetric part $T_{[\mu\nu\vrho]}$, and $(1)$ is the unique trace $T_{\mu\nu}{}^\nu$. Namely, the contraction out of two traces is $T_{\mu\nu}{}^\nu\,T^{\mu\vrho}{}_\vrho$; the contraction out of two totally antisymmetric tensors is $T^{[\mu\nu\vrho]}\,T_{[\mu\nu\vrho]}$, which is proportional to $T^{\mu\nu\vrho}\,T_{[\mu\nu\vrho]}$; the scalar contraction involving the remaining irreducible component can be replaced by $T^{\mu\nu\vrho}\,T_{\mu\nu\vrho}$; and the second contraction can be replaced by $T^{\mu\nu\vrho}\,T_{\mu\vrho\nu}$, since 
\begin{equation}
T^{\mu\nu\vrho}\,T_{\mu\vrho\nu} = \tfrac{1}{2}\,T^{\mu\nu\vrho}\,T_{\mu\nu\vrho} - \tfrac{1}{4}\,T^{\mu\nu\vrho}\,T_{[\mu\nu\vrho]}.
\end{equation}
Therefore, the three contractions in \eqref{MollerAction} can be chosen as an independent basis. 
 
Using the variations
\begin{equation}
\delta\,e = -e\,e_\mu{}^a\,\delta\,e^\mu{}_a, \quad
\delta\,T_\mu{}^\vrho{}_\nu = e_{[\mu}{}^a\,D_{\nu]}\,\delta^\vrho{}_a,
\end{equation}
one can compute the equations of motion of the tree pieces in \eqref{MollerAction} \cite{Moller1978}:
\begin{equations}
\frac{e_{\mu a}}{e}\,\frac{\delta}{\delta e^\nu{}_a}\,\int \diff^dx\,e\,T^{\alpha\beta\gamma}\,T_{\alpha\beta\gamma} =&\, - g_{\mu\nu}\,T^{\alpha\beta\gamma}\,T_{\alpha\beta\gamma} 
-2\,T_{\alpha\mu\beta}\,T^\alpha{}_\nu{}^\beta - 4\,T_{\mu\nu}{}^\alpha\,T_{\alpha\beta}{}^\beta \,+\nn\\
& + 4\,T_{\mu\alpha\beta}\,T_\nu{}^{\alpha\beta} - 4\,D_\alpha\,T_{\mu\nu}{}^\alpha,\\
\frac{e_{\mu a}}{e}\,\frac{\delta}{\delta e^\nu{}_a}\,\int \diff^dx\,e\,T^{\alpha\beta\gamma}\,T_{\alpha\gamma\beta} =&\, - g_{\mu\nu}\,T^{\alpha\beta\gamma}\,T_{\alpha\gamma\beta} 
-2\,T_{\mu\alpha\nu}\,T^{\alpha\beta}{}_\beta +2\,T_{\mu\alpha\beta}\,T_\nu{}^{\beta\alpha} \,+\nn\\
& - 2\,T_{\nu\mu\alpha}\,T^{\alpha\beta}{}_\beta -2\,D_\alpha\,T_\mu{}^\alpha{}_\nu - 2\,D_\alpha\,T_{\nu\mu}{}^\alpha,\\
\frac{e_{\mu a}}{e}\,\frac{\delta}{\delta e^\nu{}_a}\,\int \diff^dx\,e\,T^{\alpha\beta}{}_\beta\,T_{\alpha\gamma}{}^\gamma =&\, g_{\mu\nu}\,T^{\alpha\beta}{}_\beta\,T_{\alpha\gamma}{}^\gamma + 2\,g_{\mu\nu}\,D_\alpha\,T^{\alpha\beta}{}_\beta -2\,D_\nu\,T_{\mu\alpha}{}^\alpha.
\end{equations}
Therefore, the equations of motion $E_{\mu\nu} := \frac{e_{\mu a}}{e}\frac{\delta S_{\textsc{mhs}}}{\delta e^\nu{}_a} = 0$, decomposed in symmetric and antisymmetric part $E_{\mu\nu} = E_{\mu\nu}^{(\text{sym})} + E_{\mu\nu}^{\text{(anti)}}$, with $E_{\mu\nu}^{\text{(sym)}} = E_{\nu\mu}^{\text{(sym)}}$, and $E_{\mu\nu}^{\text{(anti)}} = -E_{\nu\mu}^{\text{(anti)}}$, respectively read
\begin{equations}
E_{\mu\nu}^{\text{(sym)}} =&\, 2\,\alpha_1\,T_{\mu\alpha\beta}\,T_{\nu}{}^{\alpha\beta} 
+ \alpha_2\,T_{\mu\alpha\beta}\,T_{\nu}{}^{\beta\alpha} -\alpha_1\,T_{\alpha\mu\beta}\,T^\alpha{}_{\nu}{}^\beta\,+\nn\\
& - (\alpha_1+\tfrac{\alpha_2}{2})(T_{(\mu\nu)}{}^\alpha\,T_{\alpha\beta}{}^\beta + D_\alpha\,T_{(\mu\nu)}{}^\alpha) -\tfrac{\alpha_3}{2}\,D_{(\mu}\,T_{\nu)\alpha}{}^\alpha \,+\nn\\
& - \tfrac{1}{2}\,g_{\mu\nu}\,(\alpha_1\,T^{\alpha\beta\gamma}\,T_{\alpha\beta\gamma} + \alpha_2\,T^{\alpha\beta\gamma}\,T_{\alpha\gamma\beta}-\alpha_3\,T^{\alpha\beta}{}_\beta\,T_{\alpha\gamma}{}^\gamma
-2\,\alpha_3\,D_\alpha\,T^{\alpha\beta}{}_\beta),\label{MollerEqSym}\\
E_{\mu\nu}^{\text{(anti)}} =&\,(\tfrac{\alpha_2}{2}- \alpha_1)\,(T_{[\mu\nu]}{}^\alpha\,T_{\alpha\beta}{}^\beta + D_\alpha\,T_{[\mu\nu]}{}^\alpha) 
-(\alpha_2+\tfrac{\alpha_3}{2})\,(T_{\mu\alpha\nu}\,T^{\alpha\beta}{}_\beta + D_\alpha\,T_{\mu}{}^\alpha{}_{\nu}).\label{MollerEqAnti}
\end{equations}
where in the antisymmetric part we use the trace of the Bianchi identity \eqref{TraceBianchiWeitzenbock}. Using the choice \eqref{GenRelCond2}, the antisymmetric equation is trivial (and this is the unique case) and the symmetric part, using \eqref{RicciWeitzenbock} and \eqref{RicciScalarWeitzenbock}, is recognised to be the Einstein equations. To make this manifest, use \eqref{RicciWeitzenbock} and \eqref{RicciScalarWeitzenbock} to replace $D_{(\mu}\,T_{\nu)\alpha}{}^\alpha$ and $D_\alpha\,T^{\alpha\beta}{}_\beta$:
\begin{align}
D_{(\mu}\,T_{\nu)\alpha}{}^\alpha=&\, -2\,R_{\mu\nu} - T_{\mu\alpha\beta}\,T_{\nu}{}^{\alpha\beta} - T_{\mu\alpha\beta}\,T_{\nu}{}^{\beta\alpha} \,+\nn\\
& + \tfrac{1}{2}\,T_{\alpha\mu\beta}\,T^\alpha{}_{\nu}{}^\beta + T_{(\mu\nu)}{}^\alpha\,T_{\alpha\beta}{}^\beta + D_\alpha\,T_{(\mu\nu)}{}^\alpha,\label{DTRicci}\\
D_\alpha\,T^{\alpha\beta}{}_\beta =&\, -\tfrac{1}{2}\,R-\tfrac{1}{8}\,T^{\alpha\beta\gamma}\,T_{\alpha\beta\gamma} - \tfrac{1}{4}\,T^{\alpha\beta\gamma}\,T_{\alpha\gamma\beta} - \tfrac{1}{2}\,T^{\alpha\beta}{}_\beta\,T_{\alpha\gamma}{}^\gamma.\label{DTRicciScalar}
\end{align}
Moreover, it will be useful for the following to decompose the torsion in the components in transforming in the Young tableaux $(2,1)$ and $(1,1,1)$, the first one including the trace:
\begin{equation}\label{Tdec}
T_{\mu\nu\vrho} = f_{\mu\vrho\nu} + t_{\mu\nu\vrho},
\end{equation}
where\footnote{The inverse relations read $f_{\mu\nu\vrho} = \tfrac{1}{3}\,(T_{\mu\nu\vrho} + T_{\vrho\mu\nu} - 2\,T_{\nu\vrho\mu})$ and $t_{\mu\nu\vrho} = \tfrac{1}{3}\,(T_{\mu\nu\vrho} + T_{\vrho\mu\nu} + T_{\nu\vrho\mu})$.}
\begin{equation}
f_{\mu\nu\vrho} = -f_{\nu\mu\vrho}, \quad f_{[\mu\nu\vrho]} = 0, \quad
t_{\mu\nu\vrho} = \tfrac{1}{6}\,t_{[\mu\nu\vrho]}.
\end{equation}
These symmetries imply that the following identities are satisfied:
\begin{equations}
f_{\alpha\beta\gamma}\,f^{\alpha\gamma\beta} - \tfrac{1}{2}\,f_{\alpha\beta\gamma}\,f^{\alpha\beta\gamma} = 0,\label{ftids1}\\
f^{\alpha\beta}{}_\beta\,f_{\alpha\mu\nu} - f^{\alpha\beta}{}_\beta\,f_{\mu\nu\alpha} = 0,\\
f_{\alpha\beta\gamma}\,t^{\alpha\beta\gamma} = 0,\\
f_{(\mu}{}^{\alpha\beta}\,t_{\nu)\alpha\beta} + \tfrac{1}{2}\,f^{\alpha\beta}{}_{(\mu}\,t_{\nu)\alpha\beta} = 0,\\
-\tfrac{1}{2}\,f_{\alpha\beta\mu}\,f^{\alpha\beta}{}_\nu + f_{\mu\alpha\beta}\,f_\nu{}^{\alpha\beta} - f_{\mu\alpha\beta}\,f_\nu{}^{\beta\alpha} = 0,\\
f_{\alpha\beta\mu}\,f^{\alpha\beta}{}_\nu + f_{\alpha\beta(\mu}\,f_{\nu)}{}^{\alpha\beta} = 0\label{ftids6}.
\end{equations}
Using the replacement rules \eqref{DTRicci}--\eqref{DTRicciScalar}, the decomposition of the torsion \eqref{Tdec}, and the identities \eqref{ftids1}--\eqref{ftids6} in the equations of motion \eqref{MollerEqSym} and \eqref{MollerEqAnti}, one obtains
\begin{align}
E_{\mu\nu}^{\text{(sym)}} =&\,\alpha_3\,(R_{\mu\nu}-\tfrac{1}{2}\,R\,g_{\mu\nu}) - (4\,\alpha_1-\alpha_2+\tfrac{\alpha_3}{2})\,f_{(\mu}{}^{\alpha\beta}\,t_{\nu)\alpha\beta} \,+\nn\\
& + (\alpha_1 - \alpha_2 - \tfrac{\alpha_3}{4})\,(t_{\mu\alpha\beta}\,t_\nu{}^{\alpha\beta}-\tfrac{1}{2}\,g_{\mu\nu}\,t_{\alpha\beta\gamma}\,t^{\alpha\beta\gamma}) \,+\nn\\
& + (2\,\alpha_1+\alpha_2+\alpha_3)\,(\tfrac{1}{2}\,D_\alpha\,f^\alpha{}_{(\mu\nu)} + \tfrac{1}{2}\,f^{\alpha\beta}{}_\beta\,f_{\alpha(\mu\nu)} \,+\nn\\
& + f_{\mu\alpha\beta}\,f_\nu{}^{\beta\alpha} - \tfrac{1}{4}\,g_{\mu\nu}\,f_{\alpha\beta\gamma}\,f^{\alpha\beta\gamma}),\label{MollerEomSymFinal}\\
E_{\mu\nu}^{\text{(anti)}}  =&\,\tfrac{1}{2}\,(2\,\alpha_1 + \alpha_2 + \alpha_3)\,(D_\alpha\,f^\alpha{}_{[\mu\nu]} + f^{\alpha\beta}{}_\beta\,f_{\alpha[\mu\nu]} \,+\nn\\ 
& + (-2\,\alpha_1 + 2\,\alpha_2 + \tfrac{\alpha_3}{2})\,(D_\alpha\,t^\alpha{}_{\mu\nu} + f^{\alpha\beta}{}_\beta\,t_{\alpha\mu\nu}).\label{MollerEomAntiFinal}
\end{align}
Using the choice \eqref{GenRelCond2}, all the combinations of the constants $\alpha_1, \alpha_2, \alpha_3$ vanish, and the symmetric part gives the Einstein equations.

\section{Linearisation of M{\o}ller-Hayashi-Shirafuji theory}\label{LinTorsion}

The linearisation of the M{\o}ller-Hayashi-Shirafuji theory \cite{Moller1978, Hayashi:1979qx} is obtained by defining
\begin{equation}
e_\mu{}^a = \delta_\mu^a + \vepsilon\,\overline{h}_\mu{}^a, \quad
g_{\mu\nu} = \eta_{\mu\nu} + \vepsilon\,h_{\mu\nu},
\end{equation}
where $\vepsilon$ is infinitesimal. Moreover, defining
\begin{equation}
\overline{h}_\mu{}^\nu = \overline{h}_\mu{}^a\,\delta_a^\nu, \quad
\overline{h}_{\mu\nu} = \overline{h}_\mu{}^\vrho\,\eta_{\vrho\nu} = \overline{h}_\mu{}^a\,\delta_a{}^\vrho\,\eta_{\vrho\nu}
\end{equation}
and combining the linearisation of $g_{\mu\nu}$ and $e_\mu{}^a$, we get that $h_{\mu\nu}$ is the symmetric part $\overline{h}_{\mu\nu} + \overline{h}_{\nu\mu}$. So, denoting with $b_{\mu\nu} = \overline{h}_{\mu\nu} - \overline{h}_{\nu\mu}$ the antisymmetric part, we can decompose 
\begin{equation}\label{DecHBar}
\overline{h}_{\mu\nu} = \tfrac{1}{2}\,(h_{\mu\nu} + b_{\mu\nu}).
\end{equation}
Similarly for the inverse vielbein:
\begin{equation}
e^\mu{}_a = \delta^\mu_a - \vepsilon\,\overline{h}_a{}^\mu, \quad \overline{h}_a{}^\nu\,\delta^a_\mu = \overline{h}_\mu{}^a\,\delta_a^\nu, \quad \overline{h}_b{}^\mu\,\delta^a_\nu = \overline{h}_\nu{}^a\,\delta_b^\mu,
\end{equation}
which are fixed in order to recover $e_\mu{}^a\,e^\nu{}_a = \delta_\mu^\nu$ and $e_\mu{}^a\,e^\mu{}_b = \delta^a_b$. 

The linearised equations of motion are
\begin{align}
\0{E}_{\mu\nu}^{\text{(sym)}} =&\,\alpha_3\,(\0{R}_{\mu\nu}-\tfrac{1}{2}\,\0{R}\,\eta_{\mu\nu}) + \tfrac{1}{2}\,(2\,\alpha_1+\alpha_2+\alpha_3)\,\de_\alpha\,\0{f}^\alpha{}_{(\mu\nu)},\label{MollerEomSymLin}\\
\0{E}_{\mu\nu}^{\text{(anti)}} =&\,(- 2\,\alpha_1 + 2\,\alpha_2 + \tfrac{\alpha_3}{2})\,\de_\alpha\,\0{t}^\alpha{}_{\mu\nu} + \tfrac{1}{2}\,(2\,\alpha_1 + \alpha_2 + \alpha_3)\,\de_\alpha\,\0{f}^\alpha{}_{[\mu\nu]}.\label{MollerEomAntiLin}
\end{align} 
where
\begin{align}
\0{T}_{\mu\nu\vrho} &= \0{f}_{\mu\vrho\nu} + \0{t}_{\mu\nu\vrho} = \de_{\mu}\,\overline{h}_{\vrho\nu}- \de_{\vrho}\,\overline{h}_{\vrho\nu} = \\
&= \tfrac{1}{2}\,(\de_\mu\,h_{\vrho\nu} - \de_\vrho\,h_{\mu\nu} + \de_\mu\,b_{\vrho\nu} - \de_\vrho\,b_{\mu\nu}),\\
\0{f}_{\mu\nu\vrho} &= \tfrac{1}{2}\,\de_{[\mu}\,h_{\nu]\vrho} + \tfrac{1}{6}\,(\de_{[\mu}\,b_{\nu]\vrho} + 2\,\de_\vrho\,b_{\nu\mu}),\\
\0{t}_{\mu\nu\vrho} &= \tfrac{1}{6}\,\de_{[\mu}\,b_{\vrho\nu]},\\
\0{R}_{\mu\nu} &= \tfrac{1}{2}\,\de_\vrho\,\0{T}_{(\mu\nu)}{}^\vrho - \tfrac{1}{2}\,\de_{[\mu}\,\0{T}_{\nu)\vrho}{}^\vrho = 
-\tfrac{1}{2}\,\de_\vrho\,\0{f}^\vrho{}_{(\mu\nu)} - \tfrac{1}{2}\,\de_{(\mu}\,\0{f}_{\nu)\vrho}{}^\vrho =\nn\\
&= -\tfrac{1}{2}\,\de^2\,h_{\mu\nu} + \tfrac{1}{2}\,\de^\alpha\,\de_{(\mu}\,h_{\nu)\alpha} - \tfrac{1}{2}\,\de_\mu\,\de_\nu\,h_\alpha{}^\alpha, \\
\0{R} &= -2\,\de_\mu\,\0{T}^{\mu\nu}{}_\nu = -2\,\de_\mu\,\0{f}^{\mu\nu}{}_\nu = \de_\mu\,\de_\nu\,h^{\mu\nu} - \de^2\,h_\mu{}^\mu,
\end{align}

Since the \textsc{mhs} theory is formulated in a quadratic way, unlike General Relativity in conventional formulation, the linearised equations of motion can be obtained directly by varying the action \eqref{MollerAction} with the torsion $T_{\mu\nu\vrho}$ replaced by the linearised one $\0{T}_{\mu\nu\vrho}$:
\begin{equation}\label{MollerActionLin}
\frac{1}{2}\int\diff^d\,x\,(\alpha_1\,\0{T}^{\mu\nu\vrho}\,\0{T}_{\mu\nu\vrho}+\alpha_2\,\0{T}^{\mu\nu\vrho}\,\0{T}_{\mu\vrho\nu} +\alpha_3\,\0{T}^{\mu\nu}{}_\nu\,\0{T}_{\mu\vrho}{}^\vrho).
\end{equation}
This action can be obtained working directly at the linear level in the following way. One has to look for the most general quadratic action in the derivatives of the perturbation $\overline{h}_{\mu\nu}$, invariant under linearised diffeomorphisms, which are encoded in the following \textsc{brst} rules:\footnote{They are the linearisation of the \textsc{brst} rule for the vielbein encoding the diffeomorphisms, which are generated by the Lie derivative: $s\,e_\mu{}^a = \mathcal{L}_\xi\,e_\mu{}^a = \xi^\nu\,\de_\nu\,e_\mu{}^a + e_\nu{}^a\,\de_\mu\,\xi^\nu$, $s\,\xi^\mu = \frac{1}{2}\,\mathcal{L}_\xi\,\xi^\mu = \xi^\nu\,\de_\nu\,\xi^\mu$, setting $\xi^\mu = \vepsilon\,\0{\xi}^\mu$.}
\begin{equation}\label{BRSruleLinMHS1}
s\,\overline{h}_{\mu\nu} = \de_\mu\,\0{\xi}_\nu, \quad s\,\0{\xi}^\mu = 0.
\end{equation}
or equivalently, using \eqref{DecHBar},
\begin{equation}\label{BRSruleLinMHS2}
s\,h_{\mu\nu} = \de_\mu\,\0{\xi}_\nu + \de_\nu\,\0{\xi}_\mu,\quad
s\,b_{\mu\nu} = \de_\mu\,\0{\xi}_\nu - \de_\nu\,\0{\xi}_\mu,\quad \quad s\,\0{\xi}^\mu = 0.
\end{equation}
So, one has to study the local $s$-cohomology of ghost number zero on the space of $\overline{h}_{\mu\nu}$ and its derivatives, treated as independent variables (the so-called \emph{jet space}). This is done in Appendix \ref{BRSCohomology}. The result is that the $s$-cohomology is generated by the derivatives of the antisymmetrised combination $\de_{[\mu}\,\overline{h}_{\vrho]\nu}$, which is the linearised torsion $\0{T}_{\mu\nu\vrho}$. Thus, one has to consider the most general action quadratic in $\0{T}_{\mu\nu\vrho}$, which is a combination of the three independent quadratic scalar contractions $\0{T}^{\mu\nu\vrho}\,\0{T}_{\mu\nu\vrho}$, $\0{T}^{\mu\nu\vrho}\,\0{T}_{\mu\vrho\nu}$, $\0{T}^{\mu\nu}{}_\nu\,\0{T}_{\mu\vrho}{}^\vrho$ with arbitrary coefficients, and this is the action \eqref{MollerActionLin}.

The case of General Relativity \eqref{GenRelCond2} can be obtained directly by varying the massless Fierz-Pauli action \eqref{FP}, which is the most general quadratic action in the derivatives of the symmetric part $h_{\mu\nu}$, invariant under linearised diffeomorphisms
\begin{equation}
s\,h_{\mu\nu} = \de_\mu\,\0{\xi}_\nu + \de_\nu\,\0{\xi}_\mu, \quad s\,\0{\xi}^\mu = 0.
\end{equation}
Indeed, the symmetric part is the only propagating part of $\overline{h}_{\mu\nu}$ in General Relativity. If one starts with $\overline{h}_{\mu\nu}$, the antisymmetric $b_{\mu\nu}$ can be eliminated by modifying the \textsc{brst} rules  \eqref{BRSruleLinMHS2}. One introduces a new \textsc{brst} differential operator $s_{\textsc{fp}}$, adding an antisymmetric shift in the $b_{\mu\nu}$ transformation
\begin{equation}
s_{\textsc{fp}}\,h_{\mu\nu} = \de_\mu\,\0{\xi}_\nu + \de_\nu\,\0{\xi}_\nu,\quad
s_{\textsc{fp}}\,b_{\mu\nu} = \de_\mu\,\0{\xi}_\nu - \de_\nu\,\0{\xi}_\nu + 2\,\0{\Omega}_{\nu\mu}, \quad s_{\textsc{fp}}\,\0{\xi}_\mu = 0, \quad s_{\textsc{fp}}\,\0{\Omega}_{\mu\nu} = 0,
\end{equation}
or equivalently, putting the first two transformations together,
\begin{equation}
s_{\textsc{fp}}\,\overline{h}_{\mu\nu} = \de_\mu\,\0{\xi}_\nu + \0{\Omega}_{\nu\mu}, \quad s_{\textsc{fp}}\,\0{\xi}_\mu = 0, \quad s_{\textsc{fp}}\,\0{\Omega}_{\mu\nu} = 0,
\end{equation}
where $\0{\Omega}_{\mu\nu} = -\0{\Omega}_{\nu\mu}$ is anticommuting.  Redefining $\tilde{\Omega}_{\nu\mu} =: \de_\mu\,\0{\xi}_\nu - \de_\nu\,\0{\xi}_\mu + 2\,\0{\Omega}_{\nu\mu}$, one realises that the transformation of the antisymmetric part becomes $s_{\textsc{fp}}\,b_{\mu\nu} = \tilde{\Omega}_{\mu\nu}$, $s_{\textsc{fp}}\,\tilde{\Omega}_{\mu\nu} = 0$, so that $(b_{\mu\nu},\tilde{\Omega}_{\nu\mu})$ is a trivial doublet, and the local $s_{\textsc{fp}}$-cohomology does not depend on it \cite{Brandt:1989rd, Piguet:1995er}. On the other hand, the transformation of the symmetric part is the same as before  $s_{\textsc{fp}}\,h_{\mu\nu} = s\,h_{\mu\nu}$. Therefore, the local $s_{\textsc{fp}}$-cohomology on the space of $\{h_{\mu\nu},\0{\xi}_\mu,b_{\mu\nu},\tilde{\Omega}_{\mu\nu}\}$ and their derivatives is equivalent to the local $s$-cohomology on the space of the starting field space $\{h_{\mu\nu},\0{\xi}_\mu\}$ and derivatives.

The linearised torsion $\0{T}_{\mu\nu\vrho}$ is not invariant under the modified transformations, but it transforms with respect to $\0{\Omega}_{\mu\nu}$ as a gauge two-form with a spectator index 
\begin{equation}
s_{\textsc{fp}}\,\0{T}_{\mu\nu\vrho} = \de_\mu\,\0{\Omega}_{\nu\vrho}-\de_\vrho\,\0{\Omega}_{\nu\mu}.
\end{equation}
Requiring the arbitrary combination in \eqref{MollerActionLin} to be invariant with respect to the above transformations, one selects the peculiar combination \eqref{GenRelCond2}, which indeed gives the Einstein-Hilbert action in the non-linear case. Consistently, the action \eqref{MollerActionLin} with coefficients \eqref{GenRelCond2} does not depend on the antisymmetric part $b_{\mu\nu}$ by integrating by parts.

\section{Particle spectrum in covariant fracton gauge theory}\label{ParticleFracton}

The space of the solutions of covariant fracton gauge theory $\mathscr{M}_{\text{fr}}$ (\emph{moduli space}) is the submanifold in the space of fields $\{h^{\text{(fr)}}_{\mu\nu}\}$, identified by the equations \eqref{EqFractonf} or \eqref{EqFractonf2}, modulo the gauge invariance,\footnote{The author thanks C. Imbimbo, who suggested how to expand this and the next section} which, as stated by the \textsc{brst} rule in \eqref{BRSFracton},\footnote{In this section $\lambda$ is a simple gauge parameter. We maintain the same notation of the corresponding ghost because this should not cause confusion.} amounts to linearised longitudinal diffeomorphisms, 
\begin{equation}\label{FractonGauge}
\delta_{\text{fracton}}\,h_{\mu\nu}^{\text{(fr)}} = \de_\mu\,\de_\nu\,\lambda,\;\;\forall\;\alpha, \beta.
\end{equation}
This means that two solutions $h^{\text{(fr)}}_{\mu\nu}(x)$ and $h'^{\text{(fr)}}_{\mu\nu}(x)$ of the equations of motion, which differ by a fracton gauge transformation are identified in the moduli space. 

When $\frac{\beta}{\alpha} = -2$, as in \eqref{GenRelCond1}, the theory reduced to linearised Einstein gravity, and the gauge invariance is extended to all linearised diffeomorphisms:
\begin{equation}\label{EinsteinGauge}
\delta_{\text{diff}}\,h_{\mu\nu}^{\text{(fr)}} = \de_\mu\,\xi_\nu + \de_\nu\,\xi_\mu,\;\;\text{if}\;\; \tfrac{\beta}{\alpha} = -2.
\end{equation}
There is also a case in which the fracton theory is traceless, that is, it does not depend on $h^{\text{(fr)}\vrho}{}_\vrho$, when $h_{\mu\nu}^{\text{(fr)}}$ is decomposed in its traceless part and in its trace. One can check that this happens if and only if
\begin{equation}
\tfrac{\beta}{\alpha} = -\tfrac{2}{d-1},
\end{equation}
where $\alpha$ and $\beta$ are the parameters introduced in \eqref{ActionFr}. In that case, there is an additional gauge symmetry, whose effect is to shift the trace, corresponding to a linearised Weyl scaling of $h^{\text{(fr)}}_{\mu\nu}$:
\begin{equation}\label{WeylGauge}
\delta_{\text{Weyl}}\,h^{\text{(fr)}}_{\mu\nu} = 2\,\sigma\,\eta_{\mu\nu}, \;\;\text{if}\;\; \tfrac{\beta}{\alpha} = -\tfrac{2}{d-1}.
\end{equation}

Let us find the general solution of covariant fracton gauge theory. For definiteness, we consider the four dimensional case from now on. The problem was addressed in \cite{Afxonidis:2023pdq, Afxonidis:2024tph}, where the particle content of the theory was studied: except for some special cases, the theory describes five propagating degrees of freedom, with helicities $0, \pm 1, \pm 2$ respectively. Working in Fourier space, $p^\mu$ denotes the momentum and $\tilde{h}^{\text{(fr)}}_{\mu\nu}$ the Fourier transform of the $h^{\text{(fr)}}_{\mu\nu}$. The equations \eqref{EqFractonf2} in Fourier space are:
\begin{align}
K_{\mu\nu}{}^{\vrho\sigma}(p^2;\alpha,\beta)\,\tilde{h}^{\text{(fr)}}_{\vrho\sigma} &= 2\,\alpha\,p^2\,\tilde{h}^{\text{(fr)}}_{\mu\nu} - \tfrac{1}{2}\,(2\,\alpha - \beta)\,p_\alpha\,p_{(\mu}\,\tilde{h}^{\text{(fr)}}_{\nu)}{}^\alpha \,+\nn\\
& - \beta\,p_\mu\,p_\nu\,\tilde{h}^{\text{(fr)}} + \beta\,\eta_{\mu\nu}\,(p^2\,\tilde{h}^{\text{(fr)}} - p_\alpha\,p_\beta\,\tilde{h}^{\text{(fr)}\,\alpha\beta}) = 0,\label{FractonEomsFourier}
\end{align}
where $K(p^2;\alpha,\beta)$, viewed as a $10\times 10$ matrix, identifies a system of homogeneous linear equations in the $10$ variables $\{\tilde{h}^{\text{(fr)}}_{\mu\nu}\}$. Trivial or pure-gauge solutions are those which solve the system without any condition on the momentum, and they correspond to the gauge invariance of the theory. The number of independent trivial solutions is given by the dimension of the kernel of off-shell $K(p^2;\alpha,\beta)$ (null eigenspace, whose dimension is the number of variables decreased by the matrix rank). One can show that
\begin{equation}
\#\,\text{trivial solutions} = \dim\,\text{ker}\,K(p^2 \neq 0;\alpha,\beta)=
\begin{cases}
6 &\text{if}\;\alpha=0,\\
4 &\text{if}\;\tfrac{\beta}{\alpha}=-2,\\ 
2 &\text{if}\;\tfrac{\beta}{\alpha}=-\tfrac{2}{3},\\
1 &\text{otherwise}.
\end{cases}
\end{equation}
$p_\mu\,p_\nu\,\lambda$ is always a trivial solution; in the linearised Einstein case $\frac{\beta}{\alpha}=-2$, this solution is encompassed in $p_{(\mu}\,\xi_{\nu)}$, which are four independent trivial solutions; in the traceless case $\frac{\beta}{\alpha}=-\frac{2}{3}$, besides $p_\mu\,p_\nu\,\lambda$, $2\,\eta_{\mu\nu}\,\sigma$ is trivial too. There is one peculiar case more, $\alpha = 0$, in which there is no kinetic term for the fracton gauge field, so that the theory is expected to have no propagating degrees of freedom in this case \cite{Blasi:2022mbl}.

In the massless case, the momentum can be canonically parametrised as $p^\mu = (1,0,0,1)$, such that $p^2 = 0$. The dimension of the kernel in the massless case gives the number of independent solutions \emph{without} taking into account the gauge redundancy. The result is
\begin{equation}
\dim\,\text{ker}\,K(p^2=0,\alpha,\beta) = \begin{cases} 
8 \quad \text{if}\;\frac{\beta}{\alpha}=2,\\
6 \quad\text{otherwise}.
\end{cases}
\end{equation}
The case $\frac{\beta}{\alpha}=2$ is special, because the contraction $p^\alpha\,\tilde{h}^{(\text{fr})}_{\alpha\mu}$ disappears from the equations \eqref{FractonEomsFourier}. Subtracting the number of trivial solutions to the dimension of the kernel in the various cases, one gets the number of degrees of freedom of the theory:
\begin{equation}
\#_{\text{dof}} = 
\begin{cases}
6-6 = 0, &\text{if}\;\alpha=0,\\
6-4 = 2, &\text{if}\;\tfrac{\beta}{\alpha}=-2\;\text{(linearised gravity)},\\
6-2 = 4, &\text{if}\;\tfrac{\beta}{\alpha}=-\tfrac{2}{3}\;\text{(traceless limit)},\\
8-1 = 7, &\text{if}\;\tfrac{\beta}{\alpha}=2, \\
6-1 = 5, &\text{otherwise,}
\end{cases}
\end{equation}
where we see that, as expected,  the theory has no propagating degrees of freedom when $\alpha = 0$. In the case of linearised Einstein gravity, the two degrees of freedom must describe the graviton, with helicities $\pm 2$. In the third case, we expect that the tracelessness condition eliminates a scalar particle, so that in the general case, which has five degrees of freedom, at least a scalar particle is included in the spectrum. Since the general case encompasses also the second, the graviton is also included. Only two degrees of freedom remain to be identified: they turn out to describe a particle with helicities $\pm 1$, as confirmed by the explicit solutions of the equations of motion. The two degrees of freedom more in the special case $\tfrac{\beta}{\alpha}=2$ have again helicities $\pm 1$, as we will see in the end of this section. 

Solving the equations of motion means finding a basis for the kernel in the massless case, eliminating the trivial solutions. In the general case, apart from the trivial solutions $p_\mu\,p_\nu\,\lambda$, the other five independent elements in the basis can be chosen to be equal to
\begin{equation}\label{BasisFracton}
\text{diag}\,(-1,-\tfrac{2\,\alpha + \beta}{2\,\beta},-\tfrac{2\,\alpha+\beta}{2\,\beta},1), \;\; \lambda_{\mu\nu}^{\pm 1}, \;\; \lambda_{\mu\nu}^{\pm 2}, \;\;\text{if}\;\beta \neq 0, \alpha \neq 0, \tfrac{\beta}{\alpha} = \pm 2, -\tfrac{2}{3},
\end{equation}
where $\lambda_{\mu\nu}^{s}$ are polarisation matrices with helicity $s=\pm 1, \pm 2$ (the explicit expressions are in \eqref{lambda1}--\eqref{lambda2}), and the diagonal matrix has vanishing helicity. Therefore, in the general case, covariant fracton gauge theory describes particles with helicities $0,\pm 1, \pm 2$ \cite{Afxonidis:2023pdq}. If $\beta = 0$, the scalar solution in \eqref{BasisFracton} is replaced by
\begin{equation}
\text{diag}\,(0,1,1,0),
\end{equation}
which again has vanishing helicity. If $\frac{\beta}{\alpha} = -2$, the trivial solutions $p_{(\mu}\,\xi_{\nu)}$ are completed to a basis for the kernel by $\lambda_{\mu\nu}^{\pm 2}$, as expected in linearised gravity. If $\frac{\beta}{\alpha} = -\frac{2}{3}$, the scalar solution in \eqref{BasisFracton} becomes $\eta_{\mu\nu}$, since $-\frac{2\,\alpha + \beta}{2\,\beta}\rightarrow 1$, so it is trivial, and the theory describes only the modes $\pm 1, \pm 2$. Finally, in the special case $\frac{\beta}{\alpha} = 2$, the basis has eight independent elements: one of them is the trivial one $p_\mu\,p_\nu\,\lambda$; the remaining seven ones can be chosen as
\begin{equation}
\text{diag}\,(-1,-1,-1,1), \;\;
\begin{pmatrix}
0 & -1 & \mp i & 0 \\
-1 & 0 & 0 & 0 \\
\mp i & 0 & 0 & 0 \\
0 & 0 & 0 & 0
\end{pmatrix}, \;\;
\begin{pmatrix}
0 & 0 & 0 & 0 \\
0 & 0 & 0 & -1 \\
0 & 0 & 0 & \mp i \\
0 & -1 & \mp i & 0
\end{pmatrix}, \;\; \lambda_{\mu\nu}^{\pm 2},
\end{equation}
where the diagonal solution has vanishing helicity, $\lambda_{\mu\nu}^{\pm 2}$ has helicities $\pm 2$, and the remaining two pairs of solutions have both helicities $\pm 1$. So, compared to the general case, there is one $\pm 1$ helicity particle more.

As a last observation, consider that, the polarisation matrices $\lambda_{\mu\nu}^{s}$ \eqref{lambda0}--\eqref{lambda2}, in the general solution \eqref{BasisFracton}, are transversal in the massless case, meaning that
\begin{equation}
p^\mu\,\lambda_{\mu\nu}^s = 0, \;\; s=\pm 1, \pm 2,\;\; \text{if}\; p^\mu = (1,0,0,1).
\end{equation}
Instead, the scalar mode is longitudinal 
\begin{equation}
p^\mu\,\text{diag}\,(-1,-\tfrac{2\,\alpha + \beta}{2\,\beta},-\tfrac{2\,\alpha + \beta}{2\,\beta},1)_{\mu\nu} = p_\nu.
\end{equation}
Nonetheless, this implies that the arbitrary solution $\tilde{h}_{\mu\nu}^{(\text{fr})}$ on-shell satisfies the condition
\begin{equation}\label{FractonCond}
p^\alpha\,p_{[\mu}\,\tilde{h}^{\text{(fr)}}_{\nu]\alpha} = 0 \;\;\text{if}\; p^\mu = (1,0,0,1),
\end{equation}
which will be useful in the following. 

\section{Particle spectrum of MHS theory and fracton embedding}\label{ParticleMHS}

Let us now find the particle spectrum for the \textsc{mhs} theory (see also \cite{Hayashi:1979qx}), following the same method as in the previous section. It will be convenient to redefine the constants $\alpha_1$ and $\alpha_2$ as
\begin{equation}\label{NewMHSConstants}
\tilde{\alpha}_1 = 2\,\alpha_1 + \alpha_2, \quad
\tilde{\alpha}_2 = 2\,\alpha_1 - \alpha_2.
\end{equation}
The moduli space $\mathscr{M}_{\textsc{mhs}}^{\text{(lin)}}$ of the linearised \textsc{mhs} theory is the submanifold in the field space $\{h_{\mu\nu}$, $b_{\mu\nu}\}$ identified by the equations of motion \eqref{MollerEomSymLin}--\eqref{MollerEomAntiLin}, modulo the gauge invariance, which amounts to linearised diffeomorphisms acting both on $h_{\mu\nu}$ and $b_{\mu\nu}$,\footnote{In this section $\xi_\mu$, and $\Omega_{\mu\nu}$ are simple gauge parameters. We maintain the same notation of the corresponding ghosts (omitting the dot denoting the linearisation) because this should not be cause confusion.} 
\begin{equation}\label{MHSGauge}
\delta_{\text{diff}}\,h_{\mu\nu} = \de_\mu\,\xi_\nu + \de_\nu\,\xi_\mu, \;
\delta_{\text{diff}}\,b_{\mu\nu} = \de_\mu\,\xi_\nu - \de_\nu\,\xi_\mu,\;\;\forall\;\tilde\alpha_1,\tilde\alpha_2,\alpha_3,
\end{equation}
as in \eqref{BRSruleLinMHS2}. As in covariant fracton gauge theory, there are some particular cases in which the gauge symmetry is enlarged. Linearised Einstein gravity is reproduced when the parameters are as in \eqref{GenRelCond2}, or equivalently $\tilde\alpha_2 = 0, \frac{\alpha_3}{\tilde\alpha_1}=-1$, according to \eqref{NewMHSConstants}, up to an overall constant. In this case there is no dependence on $b_{\mu\nu}$, so it can be shifted away, and this is the effect of local Lorentz invariance, as previously discussed:
\begin{equation}
\delta_{\text{Lorentz}}\,h_{\mu\nu} = 0, \;
\delta_{\text{Lorentz}}\,b_{\mu\nu} = \Omega_{\mu\nu},\;\;\text{if}\;\;\tilde\alpha_2 = 0,\;\tfrac{\alpha_3}{\tilde\alpha_1}=-1.
\end{equation}
The traceless limit in \textsc{mhs} theory is reached if and only if
\begin{equation}
\tfrac{\alpha_3}{\tilde\alpha_1}=-\tfrac{1}{d-1}.
\end{equation}
in which case there is no dependence on the trace $h^\vrho{}_\vrho$ and the gauge invariance includes a Weyl scaling:
\begin{equation}
\delta_{\text{Weyl}}\,h_{\mu\nu} = 2\,\sigma\,\eta_{\mu\nu},\;\;
\delta_{\text{Weyl}}\,b_{\mu\nu} = 0,\;\;\text{if}\;\;\tfrac{\alpha_3}{\tilde\alpha_1}=-\tfrac{1}{d-1}.
\end{equation}
A last limit case is when the theory does not depend on $h_{\mu\nu}$ at all. This happens when $\tilde\alpha_1 = \alpha_3 = 0$, in which case the symmetric equation \eqref{MollerEomSymLin} trivialises, and \textsc{mhs} theory reduces to the theory for a free two-form gauge field (Ramond-Kalb field $\textsc{rk}$ \cite{Kalb:1974yc}):
\begin{equation}
\delta_{\textsc{rk}}\,h_{\mu\nu} = \Delta_{\mu\nu}, \;
\delta_{\textsc{rk}}\,b_{\mu\nu} = \de_\mu\,\xi_\nu - \de_\nu\,\xi_\mu,\;\;\text{if}\;\;\tilde\alpha_1 = \alpha_3 = 0,
\end{equation}
where $\Delta_{\mu\nu}$ is any symmetric rank-two tensor, shifting $h_{\mu\nu}$.

In order to find the general solution of \eqref{MollerEomSymLin}--\eqref{MollerEomAntiLin}, we write them in Fourier space.  Again, we consider from now on the four-dimensional case. The equations read
\begin{equations}
&\tfrac{1}{2}\,\tilde\alpha_1\,p^2\,\tilde{h}_{\mu\nu} - \tfrac{1}{4}\,(\tilde\alpha_1 - \alpha_3)\,p_\alpha\,p_{(\mu}\,\tilde{h}_{\nu)}{}^\alpha - \tfrac{1}{2}\,\alpha_3\,p_\mu\,p_\nu\,\tilde{h} \,+\nn\\
&\quad + \tfrac{1}{2}\,\alpha_3\,\eta_{\mu\nu}\,(p^2\,\tilde{h} - p_\alpha\,p_\beta\,\tilde{h}^{\alpha\beta}) + \tfrac{1}{4}\,(\tilde\alpha_1 + \alpha_3)\,p_\alpha\,p_{(\mu}\,\tilde{b}_{\nu)}{}^\alpha = 0,\label{MHSFourier1}\\
& \tfrac{1}{2}\,\tilde\alpha_2\,p^2\,\tilde{b}_{\mu\nu} - \tfrac{1}{4}\,(\tilde\alpha_1 - 2\,\tilde\alpha_2 + \tilde\alpha_3)\,p^\alpha\,p_{[\mu}\,\tilde{b}_{\nu]\alpha} - \tfrac{1}{4}\,(\tilde\alpha_1 + \alpha_3)\,p^\alpha\,p_{[\mu}\,\tilde{h}_{\nu]\alpha} = 0.\label{MHSFourier2}
\end{equations}
This is a system of homogeneous linear equations in the $10+6$ variables $(\tilde{h}_{\mu\nu},\tilde{b}_{\mu\nu})$.

Looking at the equations alone it is possible to deduce the embedding of the fracton moduli space in the \textsc{mhs} one. Indeed, one sees that, if\footnote{In direct space, this means that the linearised torsion $\0{T}_{\mu\nu\vrho}$, corresponding to the choice \eqref{FractonId}, $\0{T}_{\mu\nu\vrho} = \tfrac{1}{2}\,f_{\mu\vrho\nu}^{\text{(fr)}}$. Moreover, there is no dependence in the projected equations on $\tilde{\alpha}_2$, and that the old constants $\alpha_1$ and $\alpha_2$ appear only in the combination $2\,\alpha_1 + \alpha_2 = \tilde{\alpha}_1$, which means that, had we varied the action with, say, $\alpha_2= 0$, we would have obtained the same result. This is not surprising, since there are only two independent quadratic scalar contractions if the totally antisymmetric part of the torsion is vanishing.}
\begin{equation}\label{FractonId}
\tilde{h}_{\mu\nu} = \tilde{h}_{\mu\nu}^{\text{(fr)}}, \quad
\tilde{b}_{\mu\nu} = 0, \quad
\tilde{\alpha}_1 = 4\,\alpha, \quad
\alpha_3 = 2\,\beta,
\end{equation}
then the first equation \eqref{MHSFourier1} reduces to the fracton equation \eqref{FractonEomsFourier}, and the second equation \eqref{MHSFourier2} reduces to the condition \eqref{FractonCond}. This shows that the fracton moduli space, if $\frac{\beta}{\alpha} \neq 2$, is contained in the subsector with vanishing $b_{\mu\nu}$ of the linearised \textsc{mhs} moduli space, upon using the identification in \eqref{FractonId},
\begin{equation}\label{Embedding}
\mathscr{M}_{\text{fr}}(\tfrac{\beta}{\alpha}\neq 2) \subseteq  {\mathscr{M}_{\textsc{mhs}}^{\text{(lin)}}}_{\big{|}_{b=0}} \subset \mathscr{M}_{\textsc{mhs}}^{\text{(lin)}}.
\end{equation}
Observe that the condition $b_{\mu\nu}= 0$ in \eqref{FractonId} breaks the gauge invariance \eqref{MHSGauge}, but it leaves a residual gauge freedom when $\xi_\mu$ is exact, that is, when it generates longitudinal diffeomorphisms $\xi_\mu \propto \de_\mu\,\lambda$. This is consistently the fracton gauge invariance \eqref{FractonGauge}.

Actually, we can prove an even stronger statement: not only the fracton moduli space is included in the $b_{\mu\nu} = 0$ subsector, but it is also isomorphic to the same subsector. To see why, let us find the most general solution of the full equations \eqref{MHSFourier1}--\eqref{MHSFourier2}. At the very end, we will see that, upon restricting in the $b_{\mu\nu} = 0$ subsector, the most general solutions coincides with the fracton one, if $\frac{\beta}{\alpha} \neq 2$. 

A $16 \times 16$ matrix $K(p;\tilde\alpha_1,\tilde\alpha_2,\alpha_3)$ can be introduced, describing the system \eqref{MHSFourier1}--\eqref{MHSFourier2} in the variables $(\tilde{h}_{\mu\nu},\tilde{b}_{\mu\nu})$, viewed as a vector with $10+6$ entries. Except for the special cases in which the gauge invariance is enlarged, the dimension of the kernel in the massless case is 10, and the number of trivial solutions is 4, encoded in $(\tilde{h}_{\mu\nu},\tilde{b}_{\mu\nu})=(p_{(\mu}\,\xi_{\nu)},p_{[\mu}\,\xi_{\nu]})$.
So, there are six degrees of freedom, which can be chosen to be equal to 
\begin{equation}
(\text{diag}\,(-1,-\tfrac{\tilde\alpha_1+\alpha_3}{2\,\alpha_3},-\tfrac{\tilde\alpha_1+\alpha_3}{2\,\alpha_3},1), O), \;\; (\lambda_{\mu\nu}^{\pm 1},O), \;\; (\lambda_{\mu\nu}^{\pm 2},O),\;\; (O,\lambda_{\mu\nu}^0),
\end{equation}
where $O$ is the $4\times 4$ null matrix, and $\lambda_{\mu\nu}^s$ are the polarisation matrices in \eqref{lambda0}--\eqref{lambda2}, with helicities $s=0,\pm 1,\pm 2$. Therefore, $\textsc{mhs}$ theory describes particles with helicities $0, \pm 1, \pm 2$, encoded in the symmetric part $\tilde{h}_{\mu\nu}$, and a scalar particle, encoded in the antisymmetric part $\tilde{b}_{\mu\nu}$ -- this is the expected result, since, as well known, a two-form gauge field in four dimensions is dual to a scalar particle. Moreover, the subsector with vanishing antisymmetric part is the same as in covariant fracton gauge theory, when the special case $\frac{\beta}{\alpha} = 2$ is excluded, since, using the identification of the constants in \eqref{FractonId}, $\frac{\tilde\alpha_1+\alpha_3}{2\alpha_3} = \frac{2\alpha + \beta}{2\beta}$, so that the first solution is the same as the first solution in \eqref{BasisFracton}. Therefore, the statement in \eqref{Embedding} holds also in the stronger version
\begin{equation}
\mathscr{M}_{\text{fr}}(\tfrac{\beta}{\alpha} \neq 2) = {\mathscr{M}_{\textsc{mhs}}^{\text{(lin)}}}_{\big{|}_{b=0}}.
\end{equation}

\section{BRST formulation}\label{BRSTorsion}

In this concluding section \textsc{brst} rules \eqref{BRSFracton} defining covariant fracton gauge theory are formulated in an equivalent way, off-shell realising the embedding of covariant fractons in linearised \textsc{mhs} theory at the level of symmetries.

Consider two doublets $(B_{\mu\nu},\tilde{b}_{\mu\nu})$ and $(r_\mu,\tilde{\xi}_\mu)$, with the following transformations
\begin{equations}
& s\,B_{\mu\nu} = \tilde{b}_{\mu\nu}, \quad s\,\tilde{b}_{\mu\nu} = 0,\\
& s\,r_\mu = \tilde{\xi}_\mu, \quad s\,\tilde{\xi}_\mu = 0,
\end{equations}
requiring that both $\tilde{b}_{\mu\nu}$ and $r_\mu$ are commuting with ghost number zero, so that $B_{\mu\nu}$ is anticommuting with ghost number $-1$ and $\tilde{\xi}_\mu$ is anticommuting with ghost number one. Then, using the doublet theorem \cite{Brandt:1989rd, Piguet:1995er}, the local $s$-cohomology on the space $\{h^{(\text{fr})}_{\mu\nu}, \lambda \}$ and derivatives is equivalent to the local $s$-cohomology on the enlarged space including the two doublets $\{h^{(\text{fr})}_{\mu\nu}, \lambda, B_{\mu\nu}, \tilde{b}_{\mu\nu}, r_\mu, \tilde{\xi}_{\mu} \}$. If we redefine $\tilde{b}_{\mu\nu}$ and $\tilde{\xi}_\mu$ in this way
\begin{equation}
\tilde{b}_{\mu\nu} =: b_{\mu\nu} - \de_\mu\,r_\nu + \de_\nu\,r_\mu, \quad 
\tilde{\xi}_\mu =: \0{\xi}_\mu - \tfrac{1}{2}\,\de_\mu\,\lambda,
\end{equation}
for some $b_{\mu\nu}$ and $\0{\xi}_\mu$, the last one being anticommuting, then the consistency of the \textsc{brst} transformations implies that 
\begin{equation}
s\,b_{\mu\nu} = \de_\mu\,\0{\xi}_\nu - \de_\nu\,\0{\xi}_\mu, \quad
s\,\0{\xi}_\mu = 0.
\end{equation}
Notice that
\begin{equation}
s\,(h^{(\text{fr})}_{\mu\nu} + \de_\mu\,r_\nu + \de_\nu\,r_\mu) = \de_\mu\,\0{\xi}_\nu + \de_\nu\,\0{\xi}_\mu
\end{equation}
is the same transformation of the linearised metric $h_{\mu\nu}$ under linearised diffeomorphisms, with ghost $\0{\xi}_\mu$. So, upon defining 
\begin{equation}
h_{\mu\nu} := h_{\mu\nu}^{\text{(fr)}} + \de_\mu\,r_\nu + \de_\nu\,r_\mu, 
\end{equation}
the \textsc{brst} rules \eqref{BRSFracton} are equivalent to the transformations of linearised \textsc{mhs} theory \eqref{BRSruleLinMHS2}, together with the transformation of a one-form field $r_\mu$, whose ghost is the fracton ghost $\lambda$, shifted by the diffeomorphism ghost $\0{\xi}_\mu$, and with the transformation of a two-form field $B_{\mu\nu}$, whose transformation is the antisymmetrised derivatives of $r_\mu$, shifted by $b_{\mu\nu}$:
\begin{equation}
\begin{cases}
s\,\lambda = 0 ,\\
s\,h_{\mu\nu}^{(\text{fr})} = \de_\mu\,\de_\nu\,\lambda
\end{cases} 
\Leftrightarrow
\begin{cases}
s\,\0{\xi}_\mu = 0, \quad s\,\lambda = 0,\\
s\,h_{\mu\nu} = \de_\mu\,\0{\xi}_\nu + \de_\nu\,\0{\xi}_\mu,\\
s\,b_{\mu\nu} = \de_\mu\,\0{\xi}_\nu - \de_\nu\,\0{\xi}_\mu,\\
s\,r_\mu = \0{\xi}_\mu - \tfrac{1}{2}\,\de_\mu\,\lambda, \\
s\,B_{\mu\nu} = b_{\mu\nu} - \de_{\mu}\,r_{\nu} + \de_{\nu}\,r_{\mu}.
\end{cases}
\end{equation}
The local $s$-cohomology on $\{h_{\mu\nu}^{\text{(fr)}},\lambda\}$ is equivalent to that on $\{B_{\mu\nu}, h_{\mu\nu}, b_{\mu\nu}, r_\mu, \0{\xi}_\mu, \lambda\}$. In other words, the doublet $(r_\mu,\tilde{\xi}_\mu)$ implements the condition ``$\0{\xi}_\mu = \tfrac{1}{2}\,\de_\mu\,\lambda$", which is implied by ``$r_\mu = 0$"; and the doublet $(B_{\mu\nu},\tilde{b}_{\mu\nu})$ implements the condition ``$b_{\mu\nu} = \de_\mu\,r_\nu - \de_\nu\,r_\mu$", which is implied by ``$B_{\mu\nu} = 0$". In this sense, the covariant fracton gauge transformations can be thought as the restriction of linearised diffeomorphisms to the longitudinal ones, as they were firstly introduced \cite{Blasi:2022mbl}, and covariant fracton gauge theory can be thought as the restriction of the linearised \textsc{mhs} theory in the case in which the antisymmetric part of the linearised vielbein is $\diff$-exact. As a consequence, the linearised torsion is cohomologically equivalent to the fracton field strength in this formulation. Indeed, the linearised vielbein reads
\begin{equation}
\overline{h}_{\mu\nu} := \tfrac{1}{2}\,(h_{\mu\nu} + b_{\mu\nu}) = \tfrac{1}{2}\,h_{\mu\nu}^{\text{(fr)}} + \de_\mu\,r_\nu + \tfrac{1}{2}\,s\,B_{\mu\nu},
\end{equation}
so that the linearised torsion becomes
\begin{equation}
\0{T}_{\mu\vrho\nu} = \de_{[\mu}\,\overline{h}_{\nu]\vrho} = \tfrac{1}{2}\,f_{\mu\nu\vrho}^{\text{(fr)}} + s\,(\tfrac{1}{2}\,\de_{[\mu}\,B_{\nu]\vrho}),
\end{equation}
which, in particular, implies that the totally antisymmetric part of the torsion is cohomologically trivial
\begin{equation}
\0{t}_{\mu\nu\vrho} = s\,(\tfrac{1}{6}\,\de_{[\mu}\,B_{\vrho\nu]}).
\end{equation}
This is the reason why the constants $\alpha_1$ and $\alpha_2$ appear only in the combination $2\,\alpha_1 + \alpha_2$ in embedding covariant fracton gauge theory in linearised \textsc{mhs} theory.

\section*{Acknowledgments}

The author is grateful to F. Fecit for encouragements and discussions, and he is deeply indebted with C. Imbimbo, who read the first draft of this work, pointing out a mistake and suggesting how improve it, stimulating the investigation in Sections \ref{ParticleFracton} and \ref{ParticleMHS}.

\appendix

\section{Weitzenb\"ock geometry within Cartan formalism}\label{SpinConnection}

In the formalism introduced by Cartan, one considers the vielbein one-form $e^a = e_\mu{}^a\,\diff x^\mu$ and the spin connection one-form $\omega^a{}_b = \omega_\mu{}^a{}_b\,\diff x^\mu$, and defines the torsion two-form $T^a = \frac{1}{2}\,T_\mu{}^a{}_\nu\,\diff x^\mu\,\diff x^\nu$ and the curvature two-form $R^a{}_b = \frac{1}{2}\,R_{\mu\nu}{}^a{}_b\,\diff x^\mu\,\diff x^\nu$ in the following way:
\begin{equations}
T^a(e,\omega) &= \diff\,e^a + \omega^a{}_b\,e^b,\label{CartanStructure1} \\
R^a{}_b(\omega) &= \diff\,\omega^a{}_b + \omega^a{}_c\,\omega^c{}_b.\label{CartanStructure2}
\end{equations}
$\omega_\mu{}^a{}_b$ serves as a frame connection (for flat indices). The basis or affine connection $\Gamma_\mu{}^\vrho{}_\nu$ (for curved indices) is related to $\omega_\mu{}^a{}_b$ by the requirement that the full covariant derivative of the vielbein, both with respect to $\omega_\mu{}^a{}_b$ and with respect to $\Gamma_\mu{}^\vrho{}_\nu$ (``second vielbein postulate"), vanishes\footnote{This is equivalent to define the covariant derivative with respect to the affine connection $\nabla_\mu(\Gamma)$ in terms of the covariant derivative with respect to the spin connection $\nabla_\mu(\omega)$ in this way: $\nabla_\mu(\Gamma)\,(e^\vrho{}_a\,V^a) = e^\vrho{}_a\,\nabla_\mu(\omega)\,V^a$, for any $V^a$. Indeed on one side $\nabla_\mu(\Gamma)\,(e^\vrho{}_a\,V^a) = \de_\mu\,V^\vrho + \Gamma_\mu{}^\vrho{}_\nu\,V^\nu$; on the other $e^\vrho{}_a\,\nabla_\mu(\omega)\,V^a = \de_\mu\,V^\vrho + e^\vrho{}_a\,(\de_\mu\,e_\nu{}^a + \omega_\mu{}^a{}_b\,e_\nu{}^b)\,V^\nu$, so that we obtain \eqref{BasisConnection0}.} \cite{VanNieuwenhuizen:1981ae}:
\begin{equation}\label{BasisConnection0}
\de_\mu\,e_\nu{}^a + \omega_\mu{}^a{}_b\,e_\nu{}^b - \Gamma_\mu{}^\vrho{}_\nu\,e_\vrho{}^a = 0,
\end{equation}
which is solved by 
\begin{equation}\label{BasisConnection}
\Gamma_\mu{}^\vrho{}_\nu(e,\omega) = e^\vrho{}_a\,\de_\mu\,e_\nu{}^a + \omega_\mu{}^a{}_b\,e^\vrho{}_a\,\,e_\nu{}^b,
\end{equation}
if the vielbein is invertible. If we define
\begin{equations}
T_{\mu}{}^\vrho{}_\nu(\Gamma) &= \Gamma_{[\mu}{}^\vrho{}_{\nu]},\label{BasisStructure1}\\
R^\vrho{}_{\sigma\mu\nu}(\Gamma) &= \de_{[\mu}\,\Gamma_{\nu]}{}^\vrho{}_\sigma + \Gamma_{[\mu}{}^\vrho{}_\alpha\,\Gamma_{\nu]}{}^\alpha{}_\sigma,\label{BasisStructure2}
\end{equations}
then they are related to the Cartan definitions \eqref{CartanStructure1}--\eqref{CartanStructure2} according to
\begin{equations}
T_{\mu}{}^\vrho{}_\nu(\Gamma(e,\omega)) &= T_\mu{}^a{}_\nu(e,\omega)\,e^\vrho{}_a,\label{AffineTorsion}\\
R^\vrho{}_{\sigma\mu\nu}(\Gamma(e,\omega)) &= 
R_{\mu\nu}{}^a{}_b(\omega)\,e^\vrho{}_a\,e_\sigma{}^b.
\end{equations}
This follows straightforwardly by replacing \eqref{BasisConnection} in  \eqref{BasisStructure1}--\eqref{BasisStructure2}.

General Relativity in usual formulation is recovered by choosing a torsionless spin connection. The spin connection appears in \eqref{CartanStructure1} algebraically, so that the torsionless condition can be solved uniquely for the spin connection in terms of the vielbein, assuming it to be invertible. Such a spin connection, which we denote with ${\omega^{(\textsc{gr})}}_\mu{}^a{}_b$, reads 
\begin{equation}\label{GRSpincConnection}
{\omega^{(\textsc{gr})}}_\mu{}^a{}_b = \tfrac{1}{2}\,e^{\nu a}\,\de_{[\mu}\,e_{\nu] b} - \tfrac{1}{2}\,e^\nu{}_b\,\de_{[\mu}\,e_{\nu]}{}^a -\tfrac{1}{2}\,e^{\nu a}\,e^\vrho{}_b\,e_\mu{}^c\,\de_{[\nu}\,e_{\vrho]c}.
\end{equation}
The corresponding affine connection, according to \eqref{BasisConnection}, which we denote by ${\Gamma^{\textsc{(gr)}}}_\mu{}^\vrho{}_\nu$, is the Levi-Civita connection: 
\begin{equation}\label{LeviCivitaConnection}
{\Gamma^{\textsc{(gr)}}}_\mu{}^\vrho{}_\nu = \Gamma_\mu{}^\vrho{}_\nu(e,\omega^{(\textsc{gr})}) = \tfrac{1}{2}\,g^{\vrho\alpha}\,(\de_\mu\,g_{\alpha\nu} 
+ \de_\nu\,g_{\mu\alpha} - \de_\alpha\,g_{\mu\nu}).
\end{equation} 
Indeed, using $g_{\mu\nu} = e_\mu{}^a\,e_{\nu a}$ and \eqref{GRSpincConnection},
\begin{align}
{\omega^{(\textsc{gr})}}_\mu{}^a{}_b\,e^\alpha{}_a\,\,e_\beta{}^b = \tfrac{1}{2}\,g^{\alpha\nu}\,(\de_\mu\,g_{\nu\beta} 
+ \de_\beta\,g_{\mu\nu} - \de_\nu\,g_{\mu\beta})  - e^\alpha{}_a\,\de_\mu\,e_\beta{}^a,
\end{align}
whence the result \eqref{LeviCivitaConnection} follows, by using \eqref{BasisConnection}. 

One could also look for a spin connection $\omega^{(\text{flat})}$ whose curvature is vanishing, instead of a torsionless spin connection. The flatness condition is uniquely solved by the Cartan-Maurer form out of an arbitrary Lorentz matrix $\lambda^a{}_b$:
\begin{equation}\label{OmegaFlat}
{\omega^{(\text{flat})}}^a{}_b = (\lambda^{-1}\,\diff\,\lambda)^a{}_b = \lambda_c{}^a\,\diff\lambda^c{}_b,
\end{equation}
The corresponding affine connection, according to \eqref{BasisConnection}, is again a Cartan-Maurer form:\footnote{The author thanks C. Imbimbo for pointing this out.}
\begin{equation}\label{FlatAffineConnection}
\Gamma_\mu{}^\vrho{}_\nu(e,\omega^{\text{(flat)}}) = e^\vrho{}_a\,\de_\mu\,e_\nu{}^a + {\omega^{(\text{flat})}}_\mu{}^a{}_b\,e^\vrho{}_a\,e_\nu{}^b = (\lambda_c{}^b\,e^\vrho{}_b)\,\de_\mu\,(\lambda^c{}_a\,e_\nu{}^a)
\end{equation}
reflecting the fact that it is flat: $R(\Gamma(e,\omega^{\text{(flat)}})=0$. So, starting from a globally defined frame $e^\mu{}_a$ in a parallelisable manifold, it is always possible to make a choice of $\lambda$ such that the affine connection above is the Weitzenb\"ock connection
\begin{equation}
W_\mu{}^\vrho{}_\nu = \Gamma_\mu{}^\vrho{}_\nu(e,\omega^{\textsc{(w)}}) = e^\vrho{}_a\,\de_\mu\,e_\nu{}^a.
\end{equation}
Instead, if we start from a frame and we identity $\Gamma(e,\omega^{\text{(flat)}})$ with the Weitzenb\"ock connection out of that frame, it means that the corresponding spin connection $\omega^{\textsc{(w)}}$ is vanishing, fixing the $\lambda$ matrix to be constant:
\begin{equation}\label{SpinConnectionW}
\omega^{(\textsc{w})} = 0.
\end{equation}
This is called \emph{Weitzenb\"ock condition}, which breaks the local Lorentz invariance, but it preserves the global Lorentz invariance.

A useful tensor is the \emph{contorsion}, which is defined as the difference between an arbitrary affine connection and the Levi-Civita connection\footnote{Using \eqref{BasisConnection}, we can also write the contorsion in terms of an arbitrary spin connection and the torsionless one: $-K_\mu{}^\vrho{}_\nu  = (\omega_\mu{}^a{}_b -{\omega^{\textsc{(gr)}}}_\mu{}^a{}_b)\,e^\vrho{}_a\,e_\nu{}^b$.}
\begin{equation}\label{Contorsion}
-K_\mu{}^\vrho{}_\nu(\Gamma) := \Gamma_\mu{}^\vrho{}_\nu - {\Gamma^{(\textsc{gr})}}_\mu{}^\vrho{}_\nu,  
\end{equation}
which implies that $K_{\mu\vrho\nu} = K_\mu{}^\sigma{}_\nu\,g_{\vrho\sigma}$ is antisymmetric in the last two indices. Notice that the antisymmetric part in the first and in the last index is the torsion \eqref{AffineTorsion}:
\begin{equation}
-K_{[\mu}{}^\vrho{}_{\nu]}(\Gamma) = T_\mu{}^\vrho{}_\nu (\Gamma).
\end{equation}
This expression can be solved for writing the contorsion in terms of the torsion
\begin{align}\label{TorsionContorsion}
K_{\mu\vrho\sigma} &= \tfrac{1}{2}\,(K_{\mu\vrho\nu}-K_{\nu\vrho\mu})+\tfrac{1}{2}\,(K_{\vrho\mu\nu}-K_{\nu\mu\vrho}) +\tfrac{1}{2}\,(K_{\vrho\nu\mu}-K_{\mu\nu\vrho}) = \nn\\
& = -\tfrac{1}{2}\,(T_{\mu\vrho\nu} + T_{\vrho\mu\nu} + T_{\vrho\nu\mu}) = -\tfrac{1}{2}\,(T_{\mu[\vrho\nu]}+T_{\vrho\mu\nu}).
\end{align}
We can relate the Levi-Civita connection, the Weitzenb\"ock connection and the Weitzenb\"ock torsion in the following way. On one side, using \eqref{Contorsion},
\begin{equation}
-K_\mu{}^\vrho{}_\nu(W) = W_\mu{}^\vrho{}_\nu - {\Gamma^{\textsc{(gr)}}}_\mu{}^\vrho{}_\nu.
\end{equation} 
On the other, using \eqref{TorsionContorsion},
\begin{equation}
-K_\mu{}^\vrho{}_\nu(W) = \tfrac{1}{2}\,(T_\mu{}^\vrho{}_\nu(W) - T_{\mu\nu}{}^\vrho(W) + T^\vrho{}_{\mu\nu}(W)).
\end{equation}
Therefore,
\begin{equation}
{\Gamma^{\textsc{(gr)}}}_\mu{}^\vrho{}_\nu  = W_\mu{}^\vrho{}_\nu  + \tfrac{1}{2}\,(T_\mu{}^\vrho{}_\nu(W) - T_{\mu\nu}{}^\vrho(W) + T^\vrho{}_{\mu\nu}(W)).
\end{equation}

\section{Computation of a BRST cohomology}\label{BRSCohomology}

Consider the following \textsc{brst} transformations:
\begin{equation}
s\,\overline{h}_{\mu\nu} = \de_\mu\,\0{\xi}_\nu, \quad
s\,\0{\xi}_\mu = 0.
\end{equation}
$(\overline{h}_{\mu\nu},\de_\mu\,\0{\xi}_\nu)$ is a doublet, so it does not belong to the \textsc{brst} cohomology. Since
\begin{equations}
& \tfrac{1}{2}\,s\,\de_{(\vrho}\,\overline{h}_{\mu)\nu} = \de_\vrho\,\de_\mu\,\0{\xi}_\nu,\\
& s\,\0{T}_{\vrho\nu\mu}  = 0,
\end{equations}
where $\0{T}_{\vrho\nu\mu} = \de_{[\vrho}\,\overline{h}_{\mu]\nu}$, then $(\tfrac{1}{2}\,\de_{(\vrho}\,\overline{h}_{\mu)\nu},\de_\vrho\,\de_\mu\,\0{\xi}_\nu)$ is a doublet, whereas $\0{T}_{\vrho\nu\mu}$ is in the cohomology. In general,
\begin{equations}
& \tfrac{1}{(n+2)!}\,s\,\de_{(\alpha_n}\,\dots\de_{\alpha_1}\,\de_{\vrho}\,\overline{h}_{\mu)\nu} = \de_{\alpha_n}\dots\de_{\alpha_1}\,\de_\vrho\,\de_\mu\,\0{\xi}_\nu,\\
& s\,\de_{\alpha_n}\,\dots\de_{\alpha_1}\,\0{T}_{\vrho\nu\mu}  = 0
\end{equations}
show that $(\tfrac{1}{(n+2)!}\,\de_{(\alpha_n}\,\dots\de_{\alpha_1}\,\de_{\vrho}\,\overline{h}_{\mu)\nu}, \de_{\alpha_n}\dots\de_{\alpha_1}\,\de_\vrho\,\de_\mu\,\0{\xi}_\nu)$ is a doublet, for each $n$. This suggests that the cohomology is generated by $\0{\xi}^\mu$, $\0{T}_{\vrho\nu\mu}$ and by the derivatives of the last one:
\begin{equation}
H(s) = \lbrace \0{\xi}^\mu, \0{T}_{\vrho\nu\mu}, \de_{\alpha_1}\,\0{T}_{\vrho\nu\mu}, \dots, \de_{\alpha_n}\dots\de_{\alpha_1}\,\0{T}_{\vrho\nu\mu}, \dots \rbrace,
\end{equation}
In order to show this, consider the decomposition $\de_\vrho\,\overline{h}_{\mu\nu}$ in irreducible components:\footnote{See note \ref{Young} for the notation of Young tableaux.}
\begin{equation}
(1) \otimes [(1) \oplus (2)_t \oplus (1,1)] =
(1) \oplus [(1) \oplus (2,1)_t \oplus (3)_t] \oplus
[(1) \oplus (2,1)_t \oplus (1,1,1)],
\end{equation}
which is realised by
\begin{equation}
\de_\vrho\,\overline{h}_{\mu\nu} = \tfrac{1}{6}\,(F_{\vrho\mu\nu} + F_{\vrho\nu\mu} + G_{\vrho\mu\nu} + B_{\nu\mu\vrho} + H_{\vrho\mu\nu}),
\end{equation}
where 
\begin{equation}
F \sim (1) \oplus (2,1)_t,\quad
G \sim (3)_t,\quad
B \sim (1) \oplus (2,1)_t,\quad
H \sim (1,1,1),
\end{equation}
and we choose $F$ and $B$ antisymmetric in the first two indices. 
Defining the symmetric and the antisymmetric part of $\overline{h}_{\mu\nu}$,
\begin{equation}
\overline{h}_{\mu\nu} = \tfrac{1}{2}\,(h_{\mu\nu} + b_{\mu\nu}), \quad\text{with}\quad
h_{\mu\nu} = \overline{h}_{\mu\nu} + \overline{h}_{\nu\mu}, \quad
b_{\mu\nu} = \overline{h}_{\mu\nu} - \overline{h}_{\nu\mu},
\end{equation}
the previous tensor can are realised by
\begin{equations}
F_{\mu\nu\vrho} &= \de_\mu\,h_{\nu\vrho} - \de_\nu\,h_{\mu\vrho},\\
G_{\mu\nu\vrho} &= \de_\mu\,h_{\nu\vrho} + \de_\nu\,h_{\vrho\mu} + \de_\vrho\,h_{\mu\nu} = \tfrac{1}{2}\,\de_{(\mu}\,h_{\nu\vrho)},\\
B_{\mu\nu\vrho} &= \de_\mu\,b_{\nu\vrho} - \de_\nu\,b_{\mu\vrho} - 2\,\de_\vrho\,b_{\mu\nu},\\
H_{\mu\nu\vrho} &= \de_\mu\,b_{\nu\vrho} + \de_\nu\,b_{\vrho\mu} + \de_\vrho\,b_{\mu\nu} = \tfrac{1}{2}\,\de_{[\mu}\,b_{\nu\vrho]}.
\end{equations}
Finally, if we decompose $\de_\mu\,\0{\xi}_\nu$,
\begin{equation}
\de_\mu\,\0{\xi}_\nu = \tfrac{1}{2}\,(S_{\nu\mu} + A_{\nu\mu}), \quad\text{with}\quad
S_{\nu\mu} = \de_\mu\,\0{\xi}_\nu + \de_\nu\,\0{\xi}_\mu, \quad
A_{\nu\mu} = \de_\mu\,\0{\xi}_\nu - \de_\nu\,\0{\xi}_\mu,
\end{equation} 
then the \textsc{brst} rules of the components of $\de_\vrho\,\overline{h}_{\mu\nu}$ read
\begin{equations}
& s\,F_{\mu\nu\vrho} = \de_\vrho\,A_{\nu\mu},\\
& s\,G_{\mu\nu\vrho} = \de_\mu\,S_{\nu\vrho} + \de_\nu\,S_{\vrho\mu} + \de_\vrho\,S_{\mu\nu},\\
& s\,B_{\mu\nu\vrho} = - 3\,\de_\vrho\,A_{\nu\mu},\\
& s\,H_{\mu\nu\vrho} = 0.
\end{equations} 
Therefore, we recognise two independent invariants
\begin{equation}
s\,(3\,F_{\mu\nu\vrho} + B_{\mu\nu\vrho}) = 0, \quad
s\,H_{\mu\nu\vrho} = 0,
\end{equation}
and this means that the $\0{T}_{\mu\nu\vrho}$ generates the cohomology, since it is a combination of the unique independent invariants:
\begin{equation}
\0{T}_{\mu\nu\vrho} = \tfrac{1}{6}\,(3\,F_{\mu\vrho\nu} + B_{\mu\vrho\nu} + 2\,H_{\mu\vrho\nu}).
\end{equation}

\section{Comparison with other conventions}\label{Conventions}

In \cite{Moller1978} the following tensors are defined:
\begin{equation}
\gamma_{\lambda\nu\mu} = -K_{\mu\lambda\nu} = e_\lambda{}^a\,\nabla_\mu\,e_{\nu a}, \quad
\vphi_\mu = \gamma^\nu{}_{\mu\nu} = -e_\lambda{}^a\,\nabla_\mu\,e_{\nu a}.
\end{equation}
The possible quadratic scalars contractions are chosen as
\begin{equations}
\vphi^\mu\,\vphi_\mu &= T^{\mu\nu}{}_\nu\,T_{\mu\vrho}{}^\vrho,\\
\gamma^{\mu\nu\vrho}\,\gamma_{\mu\nu\vrho} &= \tfrac{3}{4}\,T^{\mu\nu\vrho}\,T_{\mu\nu\vrho} + \tfrac{1}{2}\,T^{\mu\nu\vrho}\,T_{\mu\vrho\nu},\\
\gamma^{\mu\nu\vrho}\,\gamma_{\vrho\nu\mu} &= \tfrac{1}{4}\,T^{\mu\nu\vrho}\,T_{\mu\nu\vrho} + \tfrac{1}{2}\,T^{\mu\nu\vrho}\,T_{\mu\vrho\nu},
\end{equations}
so that the action
\begin{equation}
S_{\text{M{\o}ller}} = \int\diff^d x\,e\,(\beta_1\,\vphi^\mu\,\vphi_\mu + \beta_2\,\gamma^{\mu\nu\vrho}\,\gamma_{\mu\nu\vrho}  + \beta_3\,\gamma^{\mu\nu\vrho}\,\gamma_{\vrho\nu\mu})
\end{equation}
is equivalent to \eqref{MollerAction}, if
\begin{equation}
\beta_1 = \alpha_3, \quad
\beta_2 = 2\,\alpha_1-\alpha_2, \quad
\beta_3 = 3\,\alpha_2-2\,\alpha_1,
\end{equation}
and the Einstein-Hilbert action is recovered if
\begin{equation}
\beta_1 = -1, \quad \beta_2 = 0, \quad \beta_3 = 1.
\end{equation}

In \cite{Hayashi:1979qx}, the following tensors are defined in 4d:
\begin{equations}
{T^{\textsc{(hs)}}}^\lambda{}_{\mu\nu} &= T_\nu{}^\lambda{}_\mu = e^\lambda{}_a\,\de_{[\nu}\,e_{\mu]}{}^a,\\
v_\mu &= {T^{\textsc{(hs)}}}^\lambda{}_{\lambda\mu} = T_{\mu\lambda}{}^\lambda, \\
a_\mu &= \tfrac{1}{6}\,\vepsilon_{\mu\nu\vrho\sigma}\,{T^{\textsc{(hs)}}}^{\nu\vrho\sigma} = \tfrac{1}{6}\,\vepsilon_{\mu\nu\vrho\sigma}\,T^{\sigma\nu\vrho},\\
t^{\textsc{(hs)}}_{\lambda\mu\nu} &= \tfrac{1}{2}\,({T^{\textsc{(hs)}}}_{\lambda\mu\nu} + {T^{\textsc{(hs)}}}_{\mu\lambda\nu})
+\tfrac{1}{6}\,(g_{\nu\lambda}\,v_\mu + g_{\nu\mu}\,v_\lambda)
-\tfrac{1}{3}\,g_{\lambda\mu}\,v_\nu = \nn\\
&= \tfrac{1}{2}\,(T_{\nu\lambda\mu} + T_{\nu\mu\lambda}) 
+\tfrac{1}{6}\,(g_{\nu\lambda}\,T_{\mu\vrho}{}^\vrho + g_{\nu\mu}\,T_{\lambda\vrho}{}^\vrho) 
-\tfrac{1}{3}\,g_{\lambda\mu}\,T_{\nu\vrho}{}^\vrho.
\end{equations} 
$v_\mu$, $a_\mu$ and $t_{\lambda\mu\nu}$ are the irreducible components of the torsion in 4d, with $a_\mu$ (4d dual to) the totally antisymmetric part. In particular, $t_{\lambda\mu\nu}$ is traceless with vanishing totally antisymmetric part, and it is chosen to be symmetric in the first two indices. The choice for a basis for the quadratic scalar contractions of the torsion is the following:
\begin{equations}
{t^{\textsc{(hs)}}}^{\lambda\mu\nu}\,t^{\textsc{(hs)}}_{\lambda\mu\nu} &= \tfrac{1}{2}\,T^{\lambda\mu\nu}\,T_{\lambda\mu\nu} + \tfrac{1}{2}\,T^{\lambda\mu\nu}\,T_{\lambda\nu\mu} - \tfrac{1}{2}\,T^{\lambda\mu}{}_\mu\,T_{\lambda\nu}{}^\nu,\\
v^\mu\,v_\mu &= T^{\lambda\mu}{}_\mu\,T_{\lambda\nu}{}^\nu,\\
a^\mu\,a_\mu &= \tfrac{1}{18}\,T^{\lambda\mu\nu}T_{\lambda\mu\nu}+ \tfrac{1}{9}\,T^{\lambda\mu\nu}\,T_{\lambda\nu\mu},
\end{equations} 
so that the action
\begin{align}
S_{\textsc{HS}} &= \int\diff^4 x\,e\,(a_1\,{t^{\textsc{(hs)}}}^{\lambda\mu\nu}\,t^{\textsc{(hs)}}_{\lambda\mu\nu} + a_2\,v^\mu\,v_\mu + a_3\,a^\mu\,a_\mu) = \nn\\
&= \int\diff^4 x\,e\,[(c_1-\tfrac{1}{3})\,{t^{\textsc{(hs)}}}^{\lambda\mu\nu}\,t^{\textsc{(hs)}}_{\lambda\mu\nu} + (c_2+\tfrac{1}{3})\,v^\mu\,v_\mu + (c_3-\tfrac{3}{4})\,a^\mu\,a_\mu]
\end{align}
is equivalent to \eqref{MollerAction} if
\begin{equations}
a_1 &= c_1 - \tfrac{1}{3} = \tfrac{2}{3}\,(2\,\alpha_1 + \alpha_2),\\
a_2 &= c_2 + \tfrac{1}{3} = \tfrac{1}{3}\,(2\,\alpha_1 + \alpha_2 + 3\,\alpha_3),\\
a_3 &= c_3 - \tfrac{3}{4} = 6\,(\alpha_1 - \alpha_2),
\end{equations}
and the Hilbert-Einstein action is obtained if
\begin{equation}
a_1 = - \tfrac{1}{3}, \quad
a_2 = \tfrac{1}{3}, \quad
a_3 = - \tfrac{3}{4},
\end{equation}
or equivalently,
\begin{equation}
c_1 = c_2 = c_3 = 0.
\end{equation}

\section{Polarisation matrices}\label{polarisation}

See for example \cite{VanNieuwenhuizen:1981ae}. Choosing the generator of $z$-axis rotation as
\begin{equation}
J_z = \begin{pmatrix}
0 & 0 & 0 & 0 \\
0 & 0 & -i & 0 \\
0 & i & 0 & 0 \\
0 & 0 & 0 & 0
\end{pmatrix},
\end{equation}
the polarisation vectors with helicities $\pm 1$ are
\begin{equation}
\vepsilon_\mu^{\pm} = \tfrac{1}{\sqrt{2}}\,(0,1,\pm i,0),
\end{equation}
such that
\begin{equation}
J_z\,\vepsilon_\mu^{\pm} = \pm\,\vepsilon_\mu^{\pm}.
\end{equation}
Define the polarisation matrices as
\begin{equations}
\lambda_{\mu\nu}^0 = \vepsilon_{(\mu}^+\,\vepsilon_{\nu)}^- &= \begin{pmatrix}
0 & 0 & 0 & 0 \\
0 & 0 & -i & 0 \\
0 & i & 0 & 0 \\
0 & 0 & 0 & 0
\end{pmatrix}, \label{lambda0}\\
\lambda_{\mu\nu}^{\pm 1} = \tfrac{1}{\sqrt{2}}\,p_{(\mu}\,\vepsilon_{\nu)}^{\pm} &= \frac{1}{2}\begin{pmatrix}
0 & -1 & \mp i & 0 \\
-1 & 0 & 0 & 1 \\
\mp i & 0 & 0 & \pm i \\
0 & 1 & \pm i & 0 
\end{pmatrix}, \label{lambda1}\\
\lambda_{\mu\nu}^{\pm 2} = \vepsilon_{(\mu}^{\pm}\,\vepsilon_{\nu)}^{\pm} &= \frac{1}{2}\begin{pmatrix}
0 & 0 & 0 & 0 \\
0 & 1 & \pm i & 0 \\
0 & \pm i & -1 & 0 \\
0 & 0 &  0 & 0 
\end{pmatrix}.\label{lambda2}
\end{equations}
They are such that
\begin{equation}
[J_z,\lambda_{\mu\nu}^{s}] = s\,\lambda_{\mu\nu}^{s}.
\end{equation}
Two useful formulas are the following:
\begin{equation}
\begin{pmatrix}
0 & 0 & 0 & 0 \\
0 & a & b & 0 \\
0 & b & -a & 0 \\
0 & 0 & 0 & 0
\end{pmatrix} = (a-i\,b)\,\lambda^{+2} + (a + i\,b)\,\lambda^{-2}, \quad
\begin{pmatrix}
0 & -a & -b & 0 \\
-a & 0 & 0 & a \\
-b & 0 & 0 & b \\
0 & a & b & 0
\end{pmatrix} = (a-i\,b)\,\lambda^{+1} + (a + i\,b)\,\lambda^{-1}.
\end{equation}
\bibliographystyle{JHEP}
\bibliography{ir}

\end{document}